\documentclass[pra,twocolumn,nofootinbib,showkeys,preprintnumbers,amsmath,amssymb]{revtex4}
\usepackage{graphicx}
\usepackage{dcolumn}
\usepackage{bm}
\usepackage[usenames]{color}
\usepackage{color}
\usepackage{graphicx,epsfig}

\begin{document}
\title{
Landau-type sudden transitions of quantum correlations
}
\author{
Mikhail~A.~Yurischev\footnote{yur@itp.ac.ru}
}
\affiliation{\vbox{
Institute of Problems of Chemical Physics of the Russian Academy of Sciences,
142432 Chernogolovka, Moscow Region, Russia
}}

\begin{abstract}
Sudden changes of quantum correlations in the Bell-diagonal states are well-known
effects.
They occur when the set of optimal parameters that determine the quantum correlation
consists of isolated points and optimal parameters during the evolution of the system
jump from one such point to another
(e.g., the optimal measurement angle of the quantum discord changes discontinuously
from zero to $\pi/2$ or vice versa).
However, when considering more general X quantum states, we found that quantum discord
and one-way quantum work deficit can experience sudden changes of other kinds.
Namely, the optimal measurement angle may suddenly start to shift {\em continuously}
from its stationary value 0 or $\pi/2$ to an intermediate optimal measurement angle
$\vartheta\in(0,\pi/2)$.
This leads to a new behavior of quantum correlations, which is mathematically
described by the Landau phenomenological theory of second-order phase transitions.
In addition, for the one-way quantum work deficit, we found cases where the optimal
measurement angle jumps from zero to a nonzero step less than $\pi/2$, and then
continuously changes its value.
This behavior of quantum correlation is similar to a first-order phase transition in
Landau's theory.
Dependencies of quantum discord and one-way quantum work deficit near the boundaries,
which separate regions with state-dependent (variable) and state-independent
(stationary, constant) optimal measurement angles, are examined in detail on an
example of the XXZ spin model in an external field at thermal equilibrium.
\end{abstract}


\maketitle

\section{Introduction}
\label{sect:Intro}
Quantum correlations play a crucial role in quantum information science and
technologies.
Many types of these correlations have been introduced so far, and now their properties
are carefully studied both theoretically and experimentally
\cite{HHHH09,MBCPV12,AFY14}.
The most important among quantum correlations are quantum entanglement, discord,
one-way quantum work deficit and some others \cite{ABC16}.

Many quantitative measures of quantum correlations for the mixed states involve
optimization procedure for the corresponding objective function or
functional.
For example, the entanglement of formation within the entanglement-separability
paradigm \cite{W89} is defined as the minimized average entanglement over all
ensembles of pure states realizing the given state \cite{BDSW96}.
Definitions of the quantum discord and one-way quantum work deficit contain
minimizations, respectively, of conditional and post-measurement entropies over local
measurements \cite{OZ01,Z02,HHHOSSS05}, see also \cite{MBCPV12}.

\begin{figure}[b]
\begin{center}
\epsfig{file=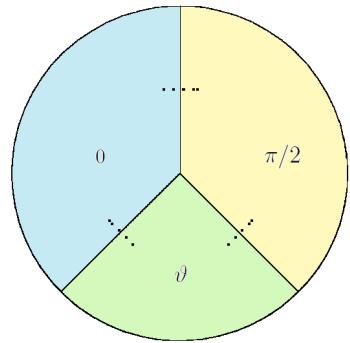,width=5.5cm}
\caption{
(Color online)
Domain of quantum discord and one-way quantum work deficit functions for the X states
(schematically).
Generally, it consists of three regions: with constant (0 and $\pi/2$) and
variable ($\vartheta$) optimal measurement angles.
Dotted lines show paths crossing the boundaries.
The transitions between 0- and $\pi/2$-regions are simple jumps of optimal parameter,
while transitions from 0- or $\pi/2$- to $\vartheta$-region correspond to the
Landau-type sudden changes of
the 
correlations.
}
\label{fig:zcirc}
\end{center}
\end{figure}

Let $\Phi(p;x)$ be a continuous and smooth objective function, where $p$ is the
vector of optimization parameters and $x$ represents additional variables, which do
not participate in the optimization process but can deform the objective function.
After optimization, one obtains the function $f(x)=\min_p\Phi(p;x)$.
It is natural to expect that the optimal parameters, $\{p_{opt}\}$, depend
continuously on the variables $x$ at least near the given optimal point.
However, it has long been noticed that the most practical constrained optimization
problems in natural sciences very often have an optimal solution at the boundaries:
``Real life optimization problems often involves one or more constraints and in most
cases, the optimal solutions to such problems lie on constraint boundaries''
\cite{RSIS09}.
If the set of optimal points is countable (for example, when the constraints are
one-dimensional segments with two endpoints each) and the above empirical
observation holds true, then the resulting function $f(x)$ will be piecewise,
consisting of analytic subfunctions (branches or phases, fractions -- in physical
language), each of which is defined in its own separate domain.
The X quantum states are a relevant example of this; see Fig.~\ref{fig:zcirc} and
details below.
%

A similar picture takes place for entangled-untangled transitions, when in the course
of evolution in time, temperature, etc., the optimal ensemble suddenly begins to
lead to a separable state and quantum entanglement disappears in a non-zero area
(volume) \cite{N98,ABV01,YE09,WHH18,SG20,Y11}.
Next, according to the Luo formula \cite{Luo08}, the quantum discord function of
Bell-diagonal states, which are a special case of X quantum states, consist of
branches that can be parametrized using the optimal measurement angles zero and
$\pi/2$ \cite{Y17}.
It is important to emphasize that the corresponding domain does not contain regions
with the variable optimal measurement angles and therefore transitions from one branch
to another are simple jumps of optimal parameter.
Such transitions have been classified as sudden changes and they are accompanied by
fractures (sharp bends) on the curves of quantum correlation dependencies
\cite{MCSV09,LC10,PKF13,CM17,BDSRSS18}.

Further development of the theory was associated with the consideration of quantum
discord in arbitrary X quantum states.
The first attempt in this direction was made by Ali et al. \cite{ARA10}.
Unfortunately, the authors decided that the extreme values of the parameters that
determine von Neumann measurements, are reached only at the endpoints.
However, shortly thereafter, counterexamples of X density matrices were given,
which clearly demonstrated the minimum of the conditional entropy function
inside the open interval, not at the endpoints of the optimization parameter segment
\cite{LMXW11,CZYYO11}.

The next step was taken in Refs.~\cite{Y14,Y14a,Y15}, where it was shown that the set
of interior optimal points form a separate continuous subregion with precise
boundaries.
Equations for such boundaries were proposed in the same papers \cite{Y14,Y14a,Y15}.
Their solutions allowed to prove that the regions under discussion are small and
narrow.
Due to the presence of such regions, we are faced with exclusion from the
above-mentioned empirical rule, since there is a return to the continuous dependence
of the optimal parameter on variables that change the state of the system and,
therefore, the corresponding objective function.

Then, similar regions with state-dependent optimal measurements were also
found in the one-way quantum work deficit \cite{YWF16,Y18,Y19}.
In Ref.~\cite{MGY19}, we proposed an optical setup scheme and performed numerical
simulations that showed that the mysterious domain with variable optimal parameter may
be detected experimentally using existing modern optical means.

This article is devoted to a detailed description of the behavior of quantum discord
and one-way quantum work deficit in the vicinity of boundaries separating the regions
with state-dependent and stationary (constant) optimal measurement angles.
We take a two-qubit XXZ model in a homogeneous magnetic field at thermal equilibrium.
This simple system already demonstrates the phenomena declared in the title of the
paper.
We show that quantum discord can undergo sudden changes that are surprisingly similar
to the Landau continuous phase transitions.
On the other hand, the behavior of the quantum work deficit is richer.
This quantum correlation under certain conditions exhibits sudden changes similar not
only to second-order phase transitions, but also to the first-order ones.

This paper is organized as follows.
The next section recalls the definitions of the one-way quantum work deficit and
quantum discord.
Model Hamiltonian and entropy expressions determining quantum correlations
are presented in Sec.~\ref{sect:Ham-rho}.
Sudden transitions of quantum work deficit and quantum discord are discussed in
Secs.~\ref{sect:w_def} and \ref{sect:QD}, respectively.
Finally, Sec.~\ref{sect:Concl} provides a brief summary and concluding remarks.

\section{Preliminaries}
\label{sect:Prelim}
Here we remember some definitions and well-known relations that will be needed in the
following sections.

Let us start with the one-way quantum work deficit (or in short, work deficit)
$\rm\Delta$, also sometimes referred to as ``thermal discord'' \cite{MBCPV12}.
For a bipartite system $AB$ in state $\rho$, it is defined as the minimum difference
between the von Neumann entropies of the post- and pre-measurement states
\cite{HHHOSSS05,SKB11,CCR15}:
\begin{equation}
   \label{eq:Delta}
   {\rm\Delta}=\min_{\{\Pi_k^B\}}S(\bar\rho)-S(\rho),
\end{equation}
where local measurements are performed for definiteness on the subsystem $B$:
$\Pi_k^B={\rm I}\otimes\Pi_k$.
In Eq.~(\ref{eq:Delta}), $\bar\rho=\sum_kp_k\rho_k$ is the weighted average of
post-measurement states
\begin{equation}
   \label{eq:rho_k}
   \rho_k=\frac{1}{p_k}({\rm I}\otimes\Pi_k)\rho({\rm I}\otimes\Pi_k)^\dagger
\end{equation}
appearing, according to the Born rule, with probabilities
$p_k={\rm Tr}(\Pi_k^B\rho\Pi_k^{B\dagger})$.
It is noteworthy that the probabilities drop out from the state after measurement:
\begin{equation}
   \label{eq:bar1_rho}
   \bar\rho=\sum_k({\rm I}\otimes\Pi_k)\rho({\rm I}\otimes\Pi_k)^\dagger
\end{equation}
what corresponds to nonselective measurement.

It is known \cite{NC10} that generalized measurements can decrease entropy, but
orthogonal projective ones do not.
Therefore, when the work deficit measures are considered, ones use the
projective measurements which satisfy to conditions $\Pi_k\Pi_i=\delta_{ik}\Pi_i$.
For a binary system, $\Pi_k=V\pi_kV^\dagger$ ($k=0,1$) with
$\pi_k=|k\rangle\langle k|$ and transformations $\{V\}$ belonging to the special
unitary group $SU_2$.
Such transformations can be parametrized by rotations
\begin{equation}
   \label{eq:V}
   V(\theta,\phi)
	 =\left(
      \begin{array}{cc}
      \cos\frac{\theta}{2}&-e^{-i\phi}\sin\frac{\theta}{2}\\
      e^{i\phi}\sin\frac{\theta}{2}&\cos\frac{\theta}{2}
      \end{array}
   \right),
\end{equation}
where $\theta\in[0,\pi]$ and $\phi\in[0,2\pi)$ are the polar and azimuthal angles,
respectively.

Similarly, the definition of quantum discord $Q$ for a composite system $AB$  can be
written as the minimum discrepancy between the two quantum versions of conditional
entropy.
On the one hand, there is a formally-defined quantum analogue of classical
conditional entropy that is given as
\begin{equation}
   \label{eq:Spre}
   S(A|B)=S(\rho)-S(\rho_B),
\end{equation}
where $\rho_B={\rm Tr}_A\rho$ is the reduced quantum state of subsystem $B$.
On the other hand, the average measurement-based quantum conditional entropy was
motivated to be given by \cite{Z00}
\begin{equation}
   \label{eq:bar_S}
   \bar S\equiv S(A|\{\Pi_k^B\})=\sum_kp_kS(\rho_{A|k}),
\end{equation}
where
\begin{equation}
   \label{eq:rho_Ak}
   \rho_{A|k}=\frac{1}{p_k}{\rm Tr}_B[({\rm I}\otimes\Pi_k)\rho({\rm I}\otimes\Pi_k)^\dagger]
\end{equation}
is the conditional state of subsystem $A$ after obtaining the outcome $k$ on the
subsystem $B$.
Now the quantum discord is written as \cite{MBCPV12,CCR15}
\begin{equation}
   \label{eq:Q}
   Q=\min_{\{\Pi_k^B\}}S(A|\{\Pi_k^B\})-S(A|B).
\end{equation}
For the quantum discord, we also restrict ourselves to the orthogonal projective
measurements (von Neumann measurements).

In the case of projective measurements, the measurement-based conditional entropy and
the entropy of post-measured state are related to each other through the condition
\cite{CRC10,H13}
\begin{equation}
   \label{eq:S-postS}
   S(A|\{\Pi_k^B\})=S(\bar\rho)-S(\bar\rho_B),
\end{equation}
where $\bar\rho$ is given by Eq.~(\ref{eq:bar1_rho}) and
${\bar\rho}_B={\rm Tr}_A{\bar\rho}$ is the state of party $B$ after the measurement.
Note that the right side of Eq.~(\ref{eq:S-postS}) is similar to the right side of
formally-defined quantum conditional entropy~(\ref{eq:Spre}).
Equation~(\ref{eq:S-postS}) allows actually compute the average measurement-based
conditional entropy.

\section{
Hamiltonian and entropy functions
}
\label{sect:Ham-rho}
Consider a system of two spins 1/2 in a uniform magnetic field $B$ with the
Hamiltonian
\begin{equation}
   \label{eq:H}
   H=-\frac{1}{2}[J(\sigma_1^x\sigma_2^x+\sigma_1^y\sigma_2^y)
	 +J_z\sigma_1^z\sigma_2^z]
	 -\frac{1}{2}B(\sigma_1^z+\sigma_2^z),
\end{equation}
where $J$ and $J_z$ are the coupling constants and $\sigma_j^x$, $\sigma_j^y$, and
$\sigma_j^z$ the Pauli spin matrices at sites $j=1,2$.
Note that there are a large number of dimeric and quasidimeric magnetic materials
similar to this system, in which quantum correlations are intensively studied
experimentally \cite{MBCPV12,AFY14}.

The energy levels of the Hamiltonian (\ref{eq:H}) are given as
\begin{equation}
   \label{eq:Ei}
   E_{1,2}=-\frac{1}{2}J_z\pm B,\qquad E_{3,4}=\frac{1}{2}J_z\pm J
\end{equation}
and hence the partition function $Z=\sum_i\exp(-E_i/T)$ equals
\begin{equation}
   \label{eq:Z}
   Z=2\biggl(e^{J_z/2T}\cosh\frac{B}{T}+e^{-J_z/2T}\cosh\frac{J}{T}\biggl),	
\end{equation}
where $T$ is the absolute temperature in energy units.
The function $Z(T,B)$ is invariant under substitutions of both $J\to-J$ and $B\to-B$.

\subsection{Pre-measurement entropies}
\label{subsect:pre}
The free energy (Helmholtz thermodynamic potential) is equal to $F(T,B)=-T\ln Z$ and,
therefore, the usual thermodynamic entropy $S=-\partial F/\partial T$ is expressed for
the system (\ref{eq:H}) by the equation
\begin{eqnarray}
   \label{eq:S_TB}
   &&S(T,B)=
	 -\frac{1}{Z}\biggl[\frac{B+J_z/2}{T}\exp\left(\frac{B+J_z/2}{T}\right)
	 \nonumber\\
	 &&-\frac{B-J_z/2}{T}\exp\left(-\frac{B-J_z/2}{T}\right)
	 \nonumber\\
   &&+\frac{J-J_z/2}{T}\exp\left(\frac{J-J_z/2}{T}\right)
	 \nonumber\\
	 &&-\frac{J+J_z/2}{T}\exp\left(-\frac{J+J_z/2}{T}\right)\biggr]+\ln Z
\end{eqnarray}
with $Z$ equals (\ref{eq:Z}).
From here one can obtain the heat capacity $C(T,B)=T\partial S/\partial T$.

The Gibbs density operator $\rho=\exp(-H/T)/Z$ has a block-diagonal structure, which is
a special case of the X-matrix:
\begin{eqnarray}
   \label{eq:rhoAB}
   \rho&=&\left(
      \begin{array}{cccc}
      a&0&0&0\\
      0&b&v&0\\
      0&v&b&0\\
      0&0&0&d
      \end{array}
   \right)
	 =\frac{1}{4}[1+s_1(\sigma_1^z+\sigma_2^z)
	 \nonumber\\
	 &&+c_1(\sigma_1^x\sigma_2^x+\sigma_1^y\sigma_2^y)+c_3\sigma_1^z\sigma_2^z],
\end{eqnarray}
where the unary and binary statistical correlations equal
\begin{equation}
   \label{eq:scc}
   s_1=a-d,\quad c_1=2v,\quad c_3=a-2b+d.
\end{equation}
The entries of matrix $\rho$ are given as
\begin{eqnarray}
   \label{eq:abdv}
   a=\frac{1}{Z}\exp[(J_z/2+B)/T],\quad
   b=\frac{1}{Z}e^{-J_z/2T}\cosh(J/T),
	 \nonumber\\
	 \\
   d=\frac{1}{Z}\exp[(J_z/2-B)/T],\quad
   v=\frac{1}{Z}e^{-J_z/2T}\sinh(J/T)
	 \nonumber
\end{eqnarray}
with the normalization condition $a+2b+d=1$.
The eigenvalues of the density matrix (\ref{eq:rhoAB}) are
\begin{equation}
   \label{eq:lam0}
   \lambda_1=a,\quad \lambda_{2,3}=b\pm v,\quad \lambda_4=d
\end{equation}
or
\begin{eqnarray}
   \label{eq:lam}
   &&\lambda_{1,4}=\frac{1}{Z}\exp[(J_z/2\pm B)/T],
	 \nonumber\\
   &&\lambda_{2,3}=\frac{1}{Z}\exp[-(J_z/2\pm J)/T].
\end{eqnarray}
In accord with (\ref{eq:lam0}), the von Neumann entropy before measurement is equal to
\begin{eqnarray}
   \label{eq:S}
	 &&S(\rho)\equiv-{\rm Tr}\rho\ln\rho=-\sum_i^4\lambda_i\ln\lambda_i
	 \nonumber\\
	 &&=-a\ln a -d\ln d
	 -(b+v)\ln(b+v)-(b-v)\ln(b-v).
	 \nonumber\\
\end{eqnarray}
(Here entropy is measured in nats; to convert this value to bits, one should divide it
by $\ln2$.)
After inserting expressions for $a$, $b$, $d$ and $v$, the relation~(\ref{eq:S})
returns to Eq.~(\ref{eq:S_TB}).

Similarly, the entropy of the reduced quantum state $\rho_B$ is given as
\begin{equation}
   \label{eq:SrhoB}
	 S(\rho_B)=-(a+b)\ln(a+b) -(b+d)\ln(b+d).
\end{equation}
Equations~(\ref{eq:S}) and (\ref{eq:SrhoB}) allow to obtain quantum pseudo-conditional
entropy (\ref{eq:Spre}).

\subsection{Post-measurement entropies}
\label{subsect:post}
Using relations from Sec.~\ref{sect:Prelim} and Eq.~(\ref{eq:rhoAB}), we get the state
of the system after measurement:
\begin{widetext}
\begin{equation}
   \label{eq:bar_rho}
   {\bar\rho}=
	 \left(
      \begin{array}{cccc}
      \frac{a+b}{2}+\frac{a-b}{2}\cos^2\!\theta&.&.&.\\ \\
      \frac{a-b}{4}e^{i\phi}\sin\!2\theta&b+\frac{a-b}{2}\sin^2\!\theta&.&.\\ \\
      \frac{v}{4}e^{i\phi}\sin\!2\theta&\frac{v}{2}\sin^2\!\theta&b-\frac{b-d}{2}\sin^2\!\theta&.\\ \\
      \frac{v}{2}e^{2i\phi}\sin^2\!\theta&\ -\frac{v}{4}e^{i\phi}\sin\!2\theta&\
			\frac{b-d}{4}e^{i\phi}\sin\!2\theta&\ \frac{b+d}{2}-\frac{b-d}{2}\cos^2\!\theta
      \end{array}
   \right)
\end{equation}
\end{widetext}
(in the upper triangular part, the dots have been put instead of the
complex-conjugated entries of this Hermitian matrix).

Using, e.g., the package Mathematica one can prove that the invariants of matrix
(\ref{eq:bar_rho}) do not contain the azimuthal angle $\phi$, therefore, the eigenvalues of
this matrix depend only on the polar angle $\theta$.
Moreover, the secular equation $\det(\bar\rho-\Lambda)= 0$ is factorized into product
of two polynomials of second degree.
This allows to get the eigenvalues of post-measurement state $\bar\rho$:
\begin{eqnarray}
   \label{eq:Lam}
	 \Lambda_{1,2}&=&\frac{1}{4}\lbrack\!\lbrack1+(a-d)\cos\theta\pm\{[a-d+(1-4b)\cos\theta]^2
	 \nonumber\\
	 &&+4v^2\sin^2\!\theta\}^{1/2}\rbrack\!\rbrack,
	 \nonumber\\ \\
	 \Lambda_{3,4}&=&\frac{1}{4}\lbrack\!\lbrack1-(a-d)\cos\theta\pm\{[a-d-(1-4b)\cos\theta]^2
	 \nonumber\\
	 &&+4v^2\sin^2\!\theta\}^{1/2}\rbrack\!\rbrack.
	 \nonumber
\end{eqnarray}
Hence, the post-measurement entropy is given as
\begin{eqnarray}
   \label{eq:tildeS_bar}
   &&\tilde S(\theta;T,B)\equiv S(\bar\rho)=2\ln2
	 \nonumber\\
	 &&-\frac{1}{4}\sum_{m,n=0}^1\biggl[1+(-1)^m(a-d)\cos\theta
	 \nonumber\\
	 && +(-1)^n\sqrt{[a-d+(-1)^m(1-4b)\cos\theta]^2+4v^2\sin^2\theta}\biggr]
	 \nonumber\\
	 &&\times\ln\biggl[1+(-1)^m(a-d)\cos\theta
	 \nonumber\\
	 &&+(-1)^n\sqrt{[a-d
	 +(-1)^m(1-4b)\cos\theta]^2+4v^2\sin^2\theta}\biggr].
	 \nonumber\\
\end{eqnarray}
The state of subsystem $B$ after the measurement is equal to
\begin{widetext}
\begin{equation}
   \label{eq:bar_rhoB}
   {\bar\rho_B}=
	 \left(
      \begin{array}{cc}
      \frac{1}{2}(1+\cos^2\theta)a+b+\frac{1}{2}d\sin^2\theta&\frac{1}{4}e^{-i\phi}\sin 2\theta\\ \\
			\frac{1}{4}e^{i\phi}\sin2\theta&\frac{1}{2}a\sin^2\theta+b+\frac{1}{2}(1+\cos^2\theta)d
      \end{array}
   \right).
\end{equation}
\end{widetext}
Entropy of this state is given as
\begin{eqnarray}
   \label{eq:Sbar_rhoB}
   S(\bar\rho_B)&=&\ln2-\frac{1}{2}\{[1+(a-d)\cos\theta]\ln[1+(a-d)\cos\theta]
	 \nonumber\\
	 &&+[1-(a-d)\cos\theta]\ln[1-(a-d)\cos\theta]\},
\end{eqnarray}
which also does not depend on $\phi$.

Both entropies $S(\bar\rho)$ and $S(\bar\rho_B)$ as functions of the argument $\theta$
are invariant under the transformation $\theta\to\pi-\theta$ and therefore this
variable can be limited to the values of $\theta\in[0,\pi/2]$.
It is easy to check by direct calculations that
\begin{equation}
   \label{eq:Sbar_rho-B}
   \frac{\partial S(\bar\rho)}{\partial\theta}\Bigg|_{\theta=0,\pi/2}=0,\qquad
   \frac{\partial S(\bar\rho_B)}{\partial\theta}\Bigg|_{\theta=0,\pi/2}=0
\end{equation}
for any $a$, $b$, $d$, and $v$.

\subsection{Conditional entropy}
\label{subsect:cond_entr}
Equations~(\ref{eq:bar_S}), (\ref{eq:S-postS}), (\ref{eq:tildeS_bar}), and
(\ref{eq:Sbar_rhoB}) allow to get the conditional entropy function
\begin{eqnarray}
   \label{eq:Sbar_rho}
   \bar S(\theta;T,B)&=&\tilde S(\theta;T,B)
	 \nonumber\\
	 &&+\frac{1+(a-d)\cos\theta}{2}\ln\frac{1+(a-d)\cos\theta}{2}
	 \nonumber\\
	 &&+\frac{1-(a-d)\cos\theta}{2}\ln\frac{1-(a-d)\cos\theta}{2}
	 \nonumber\\
\end{eqnarray}
with quantities $a$ and $d$ equal to the relations~(\ref{eq:abdv}).

Finally, according to Eq.~(\ref{eq:Sbar_rho-B}), the first derivatives of the
post-measurement and conditional entropies with respect to the measurement angle
$\theta$ vanish at boundaries,
\begin{equation}
   \label{eq:StildeSbar_rho}
   \tilde S^\prime(\theta;T,B)|_{\theta=0,\pi/2}=0,\quad
   \bar S^\prime(\theta;T,B)|_{\theta=0,\pi/2}=0,
\end{equation}
and therefore both endpoints $\theta=0$ and $\pi/2$ are stationary (critical) and
candidates for the global extrema.

So, the necessary equations are written down, and now we can proceed directly to the
study of the behavior of quantum correlations.

\begin{figure*}
\begin{center}
\epsfig{file=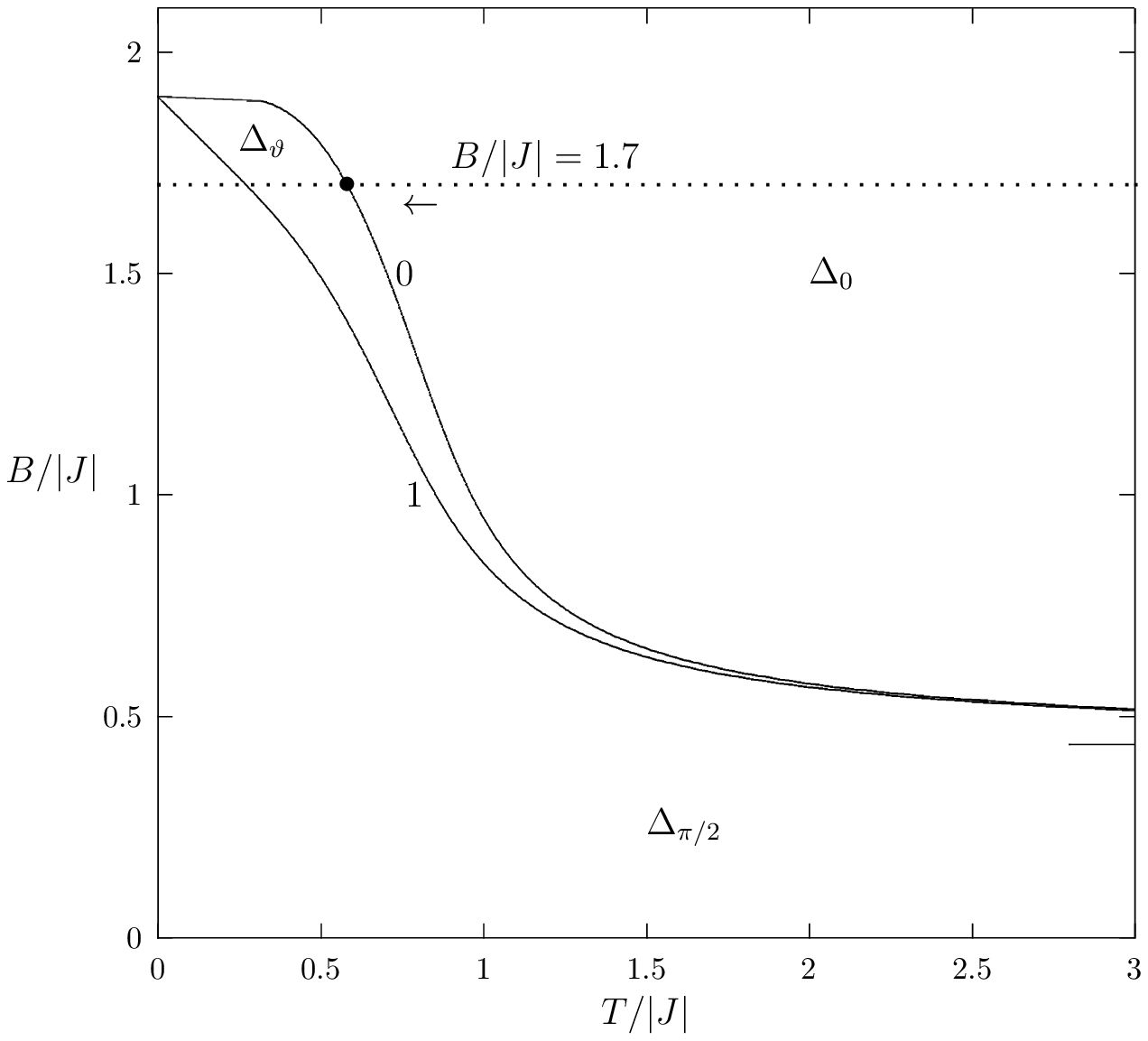,width=7.5cm}
\hspace{5mm}
\epsfig{file=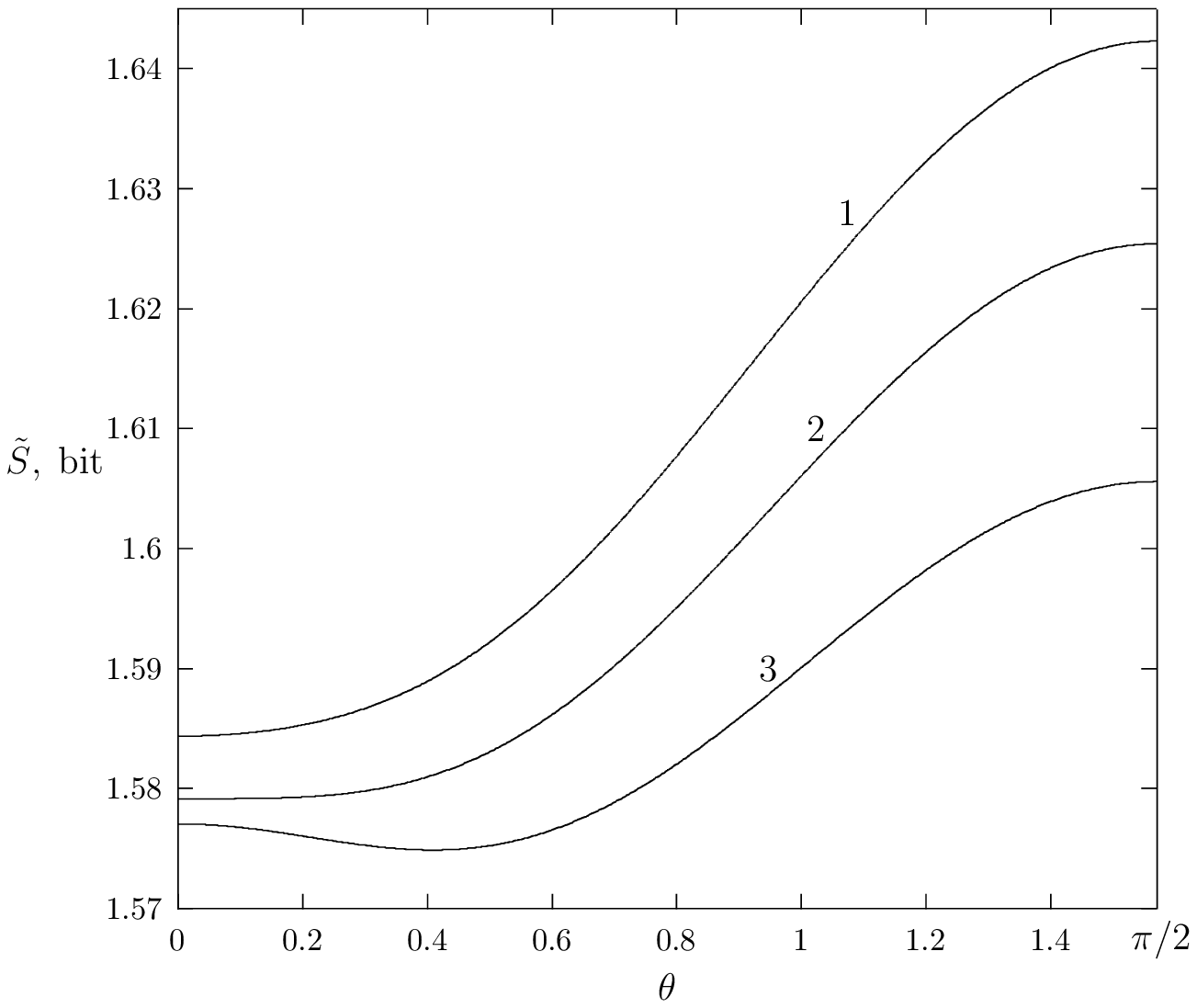,width=7.5cm}
\caption{
(left panel)
Temperature-field phase diagram of work deficit for the model with $J_z/|J|=-0.9$.
Solid lines 0 and 1 are the 0- and $\pi/2$-boundaries, respectively.
Dotted straight line is the path of system evolution.
The arrow shows the direction of evolution along the path.
Black circle ($\bullet$) has coordinates $(0.58264,1.7)$ and marks a crossing point of
the path and 0-boundary curve.
A long horizontal bar on the right ordinate axis marks the level that corresponds to
the asymptote $B/|J|=0.43589$.
(right panel)
Post-measurement entropy $\tilde S$ vs measurement angle $\theta$ for $B/|J|=1.7$ and
$T/|J|=0.65~(1), 0.5826437~(2), 0.5~(3)$.
The curve 3 has an interior minimum $\tilde S_{min}=1.57487$~bit at
$\theta_{min}=0.40953\approx23.5^\circ$.
}
\label{fig:zpdzs1way1a}
\end{center}
\end{figure*}
%
\section{Sudden transitions of work deficit}
\label{sect:w_def}
As applied to the model under discussion, the one-way quantum work deficit is given as
\begin{equation}
   \label{eq:Delt}
   {\rm\Delta}(T,B)=\min_\theta\Delta(\theta; T, B)
\end{equation}
with non-optimized work deficit function
\begin{equation}
   \label{eq:Del_th}
   \Delta(\theta;T,B)=\tilde S(\theta; T, B)-S(T, B),
\end{equation}
where $\tilde S(\theta;T, B)$ is expressed by Eq.~(\ref{eq:tildeS_bar}) and $S(T,B)$
is given by (\ref{eq:S_TB});
here $\theta$ is the current non-optimized measurement angle (the optimal measurement
angle is denoted as $\vartheta$).
Note that the derivative of the work deficit with respect to temperature is
proportional to the difference between heat capacity after ($\tilde C$) and before
($C$) optimal measurement:
\begin{equation}
   \label{eq:Var_C}
   T\frac{\partial{\rm\Delta}}{\partial T}=\tilde C(T,B)-C(T,B).
\end{equation}
This relationship points to a way to experimentally measure work deficit.
Indeed, it is necessary to measure the specific heat before and after the measurement,
and then integrate their difference normalized to temperature.

Using Eq.~(\ref{eq:Del_th}) we get the value of work deficit at the endpoint
$\theta=0$:
\begin{eqnarray}
   \label{eq:Del0a}
   &&{\rm\Delta}_0(T,B)\equiv\Delta(0;T,B)
	 \nonumber\\
	 &&=\frac{2}{Z}\Biggl[\frac{J}{T}\sinh\frac{J}{T}
	 -\ln\Biggl(\cosh\frac{J}{T}\Biggr)\cosh\frac{J}{T}\Biggr]e^{-J_z/2T}.
	 \nonumber\\
\end{eqnarray}
Its high-temperature asymptotics is
\begin{equation}
   \label{eq:Del0as1}
   {\rm\Delta}_0(T,B)=\frac{1}{4}\frac{J^2}{T^2}-\frac{1}{8}\frac{J_zJ^2}{T^3}
	 -\frac{1}{16}\frac{J^2(B^2+J^2)}{T^4}+O\Biggl(\frac{1}{T^5}\Biggr).
\end{equation}
At the second endpoint $\theta=\pi/2$, work deficit is equal to
\begin{eqnarray}
   \label{eq:Del1}
   &&{\rm\Delta}_{\pi/2}(T,B)\equiv\Delta(\pi/2;T,B)=\ln2
	 -\frac{1+r}{2}\ln\frac{1+r}{2}
	 \nonumber\\
	 &&-\frac{1-r}{2}\ln\frac{1-r}{2}-S(T,B),
\end{eqnarray}
where
\begin{equation}
   \label{eq:r}
   r=\frac{2}{Z}\Bigg[e^{J_z/T}\sinh^2\frac{B}{T}+e^{-J_z/T}\sinh^2\frac{J}{T}\Bigg]^{1/2}.
\end{equation}
High-temperature expansion reads
\begin{equation}
   \label{eq:Del1as1}
   {\rm\Delta}_{\pi/2}(T,B)=\frac{1}{8}\frac{J^2+J_z^2+B^2}{T^2}
	 +\frac{1}{8}\frac{J_z(B^2-J^2)}{T^3}+O\Biggl(\frac{1}{T^4}\Biggr).
\end{equation}
Comparing the main terms of the expansions (\ref{eq:Del0as1}) and (\ref{eq:Del1as1}),
we conclude that for $T\to\infty$ the deficit will be determined by the
${\rm\Delta}_0$-branch, if $J_z^2+B^2>J^2$, and by ${\rm\Delta}_{\pi/2}$, if
$J_z^2+B^2<J^2$.

It is clear that the one-way quantum work deficit (\ref{eq:Delt}) can be written as a
choice of three alternative branches (phases):
\begin{equation}
   \label{eq:optDelt}
   {\rm\Delta}=\min\{{\rm\Delta}_0,{\rm\Delta}_{\pi/2},{\rm\Delta}_\vartheta\},
\end{equation}
where ${\rm\Delta}_0$ and ${\rm\Delta}_{\pi/2}$ are given by
Eqs.~(\ref{eq:Del0a}) and (\ref{eq:Del1}) respectively, while ${\rm\Delta}_\vartheta$
is the amount of work deficit at the internal optimal point $\vartheta\in(0,\pi/2)$,
if, of course, it exists there.

To further investigate the problem, we will use phase diagrams for the model that
were obtained in the three-dimensional space $(s_1,c_1,c_3)$ \cite{Y19} and in the
plane $(T,B)$ for different values of $J$ and $J_z$ \cite{Y20}.
Phase diagrams consist of two regions with fixed optimal measurement angles (0 and
$\pi/2$) and one region with variable, state-dependent optimal measurement angle
$\vartheta$.
Boundary between ${\rm\Delta}_0$ and ${\rm\Delta}_{\pi/2}$ phases is given by equation
${\rm\Delta}_0={\rm\Delta}_{\pi/2}$.
Crossing this boundary leads to the usual sudden changes that were mentioned in
Introduction and which are not discussed in our article.
We are interested in transitions to the region ${\rm\Delta}_\vartheta$.
For the work deficit, as announced above, they can be of two types: (i)~continuous and
(ii)~a relatively small ($<\pi/2$) jump of the optimal measurement angle and then
continuous change of $\vartheta$.
We will study these two cases separately.

\subsection{
Continuous transitions
}
\label{subsect:2nd_order}
In the case under question, the post-measurement entropy $\tilde S(\theta)$ [and
therefore the unoptimized work deficit $\Delta(\theta)$] has one internal minimum
-- inside the interval $(0,\pi/2) $.
The 0- and $\pi/2$-boundaries of the ${\rm\Delta}_\vartheta$ region are determined by
the conditions, where the second derivatives of $\Delta(\theta)$ or, the same,
$\tilde S(\theta)$ with respect to the measurement angle $\theta$ vanish,
respectively, at $\theta=0$ and $\pi/2$ \cite {Y18}:
\begin{equation}
   \label{eq:D11}
   \Delta^{\prime\prime}(0;T,B)=0\quad
   {\rm and}\quad
   \Delta^{\prime\prime}(\pi/2;T,B)=0
\end{equation}
or
\begin{equation}
   \label{eq:S11}
   \tilde S^{\prime\prime}(0;T,B)=0\quad
   {\rm and}\quad
   \tilde S^{\prime\prime}(\pi/2;T,B)=0.
\end{equation}
These equations are based on the observation that the interior minimum of unimodal
differentiable function in the closed interval with the zero first derivatives at
endpoints appears or annihilates at the boundaries.

For the second derivatives at endpoints,
$\tilde S^{\prime\prime}_0(T,B)\equiv\tilde S^{\prime\prime}(0;T,B)$ and
$\tilde S^{\prime\prime}_{\pi/2}(T,B)\equiv\tilde S^{\prime\prime}(\pi/2;T,B)$,
calculations yield
\begin{widetext}
\begin{eqnarray}
   \label{eq:Sii0}
   {\tilde S}_0^{\prime\prime}(T,B)&=&\frac{1}{Z}\bigg\lbrack\!\bigg\lbrack\bigg\lbrace\frac{B}{T}\sinh\frac{B}{T}
	 +\biggl(\cosh\frac{B}{T}-e^{-J_z/T}\cosh\frac{J}{T}\biggr)\bigg\lbrack\frac{J_z}{T}-\ln\biggl(\cosh\frac{J}{T}\biggr)\bigg\rbrack\bigg\rbrace e^{J_z/2T}
	 \nonumber\\
   &&-\frac{1}{2}e^{-J_z/2T}\bigg\lbrack\frac{(J_z+B)/T-\ln\bigl(\cosh(J/T)\bigr)}{\exp\bigl((J_z+B)/T\bigr)-\cosh(J/B)}
	 +\frac{(J_z-B)/T-\ln\bigl(\cosh(J/T)\bigr)}{\exp\bigl((J_z-B)/T\bigr)-\cosh(J/B)}\bigg\rbrack\sinh^2\frac{J}{T}\bigg\rbrack\!\bigg\rbrack
\end{eqnarray}
\end{widetext}
and
\begin{eqnarray}
   \label{eq:Sii1}
   &&\tilde S^{\prime\prime}_{\pi/2}(T,B)=8v^2\frac{ab+bd-ad-b^2+v^2}{r^3}\ln\frac{1+r}{1-r}
	 \nonumber\\
	 &&-\frac{1}{2}(a-d)^2\left\{\frac{[1+(1-4b)/r]^2}{1+r}+\frac{[1-(1-4b)/r]^2}{1-r}\right\}\!\!,\
	 \nonumber\\
\end{eqnarray}
where $r$ is given, as before, by Eq.~(\ref{eq:r}).

High-temperature expansions of functions (\ref{eq:Sii0}) and (\ref{eq:Sii1}) are
given as
\begin{equation}
   \label{eq:hTSpmII}
   {\tilde S}_0^{\prime\prime}(T,B)|_{T\to\infty}=\frac{J_z^2-J^2+B^2}{4T^2}+\frac{J_zB^2}{4T^3}+O(1/T^4)
\end{equation}
and
\begin{equation}
   \label{eq:hTSpmII1}
   {\tilde S}_{\pi/2}^{\prime\prime}(T,B)|_{T\to\infty}=\frac{J^2-J_z^2-B^2}{4T^2}-\frac{J_zB^2}{4T^3}+O(1/T^4).
\end{equation}
Equating to zero the leading terms of these expansions, we get, in accord with
(\ref{eq:S11}), the equation of the asymptote for both boundaries:
\begin{equation}
   \label{eq:B_as}
   B/|J|=\sqrt{1-(J_z/J)^2}.
\end{equation}
This result is useful for numerical estimates.

As an example, consider a system with coupling constants $J=\pm1$ and $J_z=-0.9$.
Using Eqs.~(\ref{eq:S11})-(\ref{eq:Sii1}) we find the 0- and $\pi/2$-boundaries.
The corresponding phase diagram in the plane $(T,B)$ is shown in
Fig.~\ref{fig:zpdzs1way1a}, left.
When $T\to\infty$, curves 0 and 1 tend as predicted Eq.~(\ref{eq:B_as}) to the
horizontal asymptote $B/|J|=0.43589$.

Consider the path $B/|J|=1.7$.
It intersects the 0- and $\pi/2$-boundaries respectively at the points
$T_0/|J|=0.58264$ and $T_{\pi/2}/|J|=0.27228$ (see again Fig.~\ref{fig:zpdzs1way1a},
left).
This allows us to estimate the size (width) of the region $\Delta_\vartheta$ as
$(T_0-T_{\pi/2})/[(T_0+T_{\pi/2})/2]=0.72612\approx72.6\%$.
Thus, in the temperature-field coordinates, the width of the region with a variable
angle of optimal measurement is large.
This contrasts with the sizes of ${\rm\Delta}_\vartheta$-regions in the space
$(s_1,c_1,c_3)$, where they are tiny and narrow \cite{Y19}.

The behavior of post-measured entropy $\tilde S$ versus the measurement angle $\theta$
is shown in Fig.~\ref{fig:zpdzs1way1a}, right.
When temperatures are high enough, the curves have a minimum at the endpoint
$\theta=0$.
However, as the system cools, this minimum flattens, its behavior changes from law
$\sim \theta^2$ to $\sim \theta^4$ at $T_0/|J|=0.5826437$ (curve 2).
With a further decrease in temperature, an internal minimum appears on the curve
$\tilde S(\theta)$, and the spin system continuously enters in the region
${\rm\Delta}_\vartheta$ with the variable optimal measurement angle $\vartheta>0$.
This behavior is similar to the continuous phase transition of Landau's theory
\cite{L37,LL_StPh}.

The function $\tilde S(\theta)$ is even and its expansion in a Tailor series at
$\theta=0$ has the form
\begin{equation}
   \label{eq:S_thet}
   \tilde S(\theta; T,B)=\tilde S_0(T,B)
	 +\frac{1}{2!}\tilde S^{\prime\prime}_0(T,B)\!\cdot\!\theta^2
	 +\frac{1}{4!}\tilde S^{\prime\prime\prime\prime}_0(T,B)\!\cdot\!\theta^4+\ldots,
\end{equation}
where $\tilde S_0(T,B)$ equals $\tilde S(0;T,B)$, $\tilde S^{\prime\prime}_0(T,B)$ is
given by Eq.~(\ref{eq:Sii0}),  and $\tilde S^{\prime\prime\prime\prime}_0(T,B)$ is the
fourth derivative of post-measurement entropy with respect to the measurement angle at
$\theta=0$.
Hence the expansion of work deficit has the form
\begin{equation}
   \label{eq:D_thet}
   \tilde \Delta(\theta; T,B)={\rm\Delta}_0(T,B)
	 +\frac{1}{2!}\tilde S^{\prime\prime}_0(T,B)\!\cdot\!\theta^2
	 +\frac{1}{4!}\tilde S^{\prime\prime\prime\prime}_0(T,B)\!\cdot\!\theta^4+\ldots,
\end{equation}
where ${\rm\Delta}_0(T,B)$ is given by Eq.~(\ref{eq:Del0a}).
These series are similar to the Landau expansion of the thermodynamic potential
(Gibbs free energy) in his theory of second-order phase transitions
\cite{L37,LL_StPh}.
In our problem, $\tilde S(\theta)$ or $\Delta(\theta)$ play a role of the
thermodynamic potential and the optimal measurement angle $\vartheta$ is a 
counterpart of Landau's order parameter.
The vanishing of the coefficient at the quadratic term corresponds to the sudden
transition point.

\begin{figure}
\begin{center}
\epsfig{file=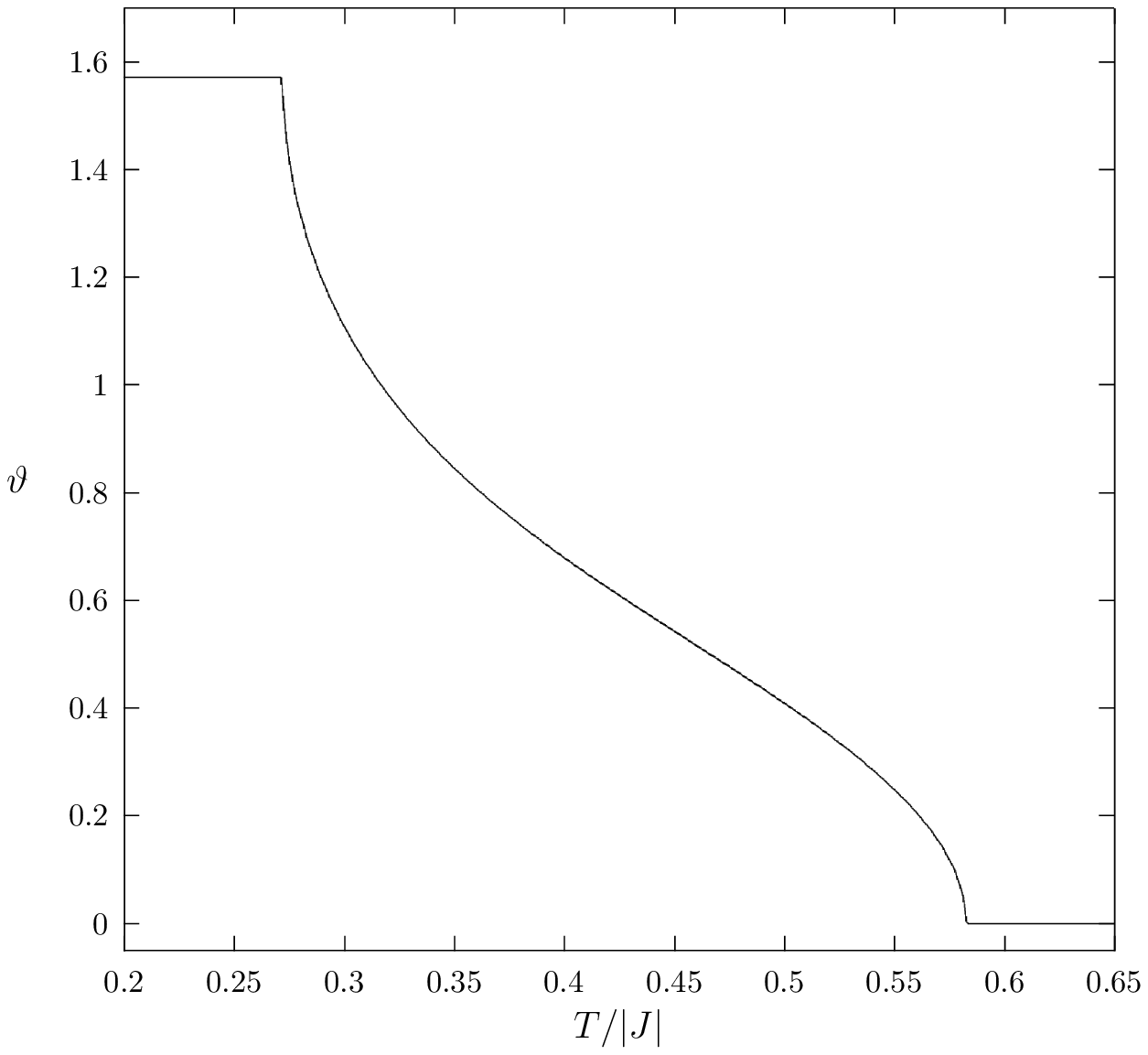,width=5.8cm}
\epsfig{file=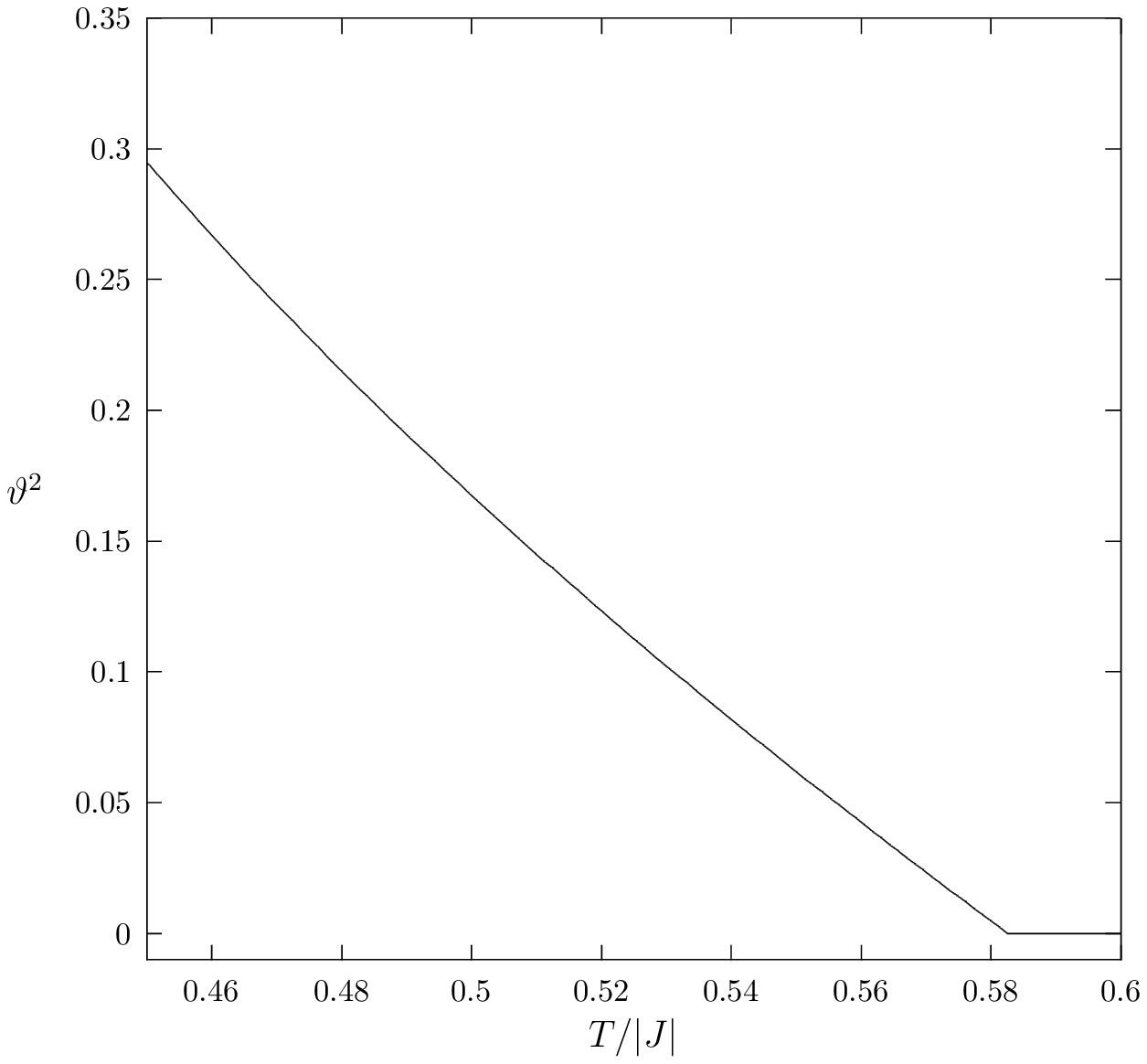,width=5.8cm}
\caption{
Optimal measurement angle $\vartheta$ (top) and its square (bottom) as functions of
temperature by  $J_z/|J|=-0.9$ and $B/|J|=1.7$.
}
\label{fig:zt209B17}
\end{center}
\end{figure}

The temperature dependence of $\vartheta$ is shown in Fig.~\ref{fig:zt209B17}, top.
%
%
The optimal measurement angle takes constant values $\vartheta=0$ or $\vartheta=\pi/2$
for $T\ge T_0=0.58264|J|$ and $T\le T_{\pi/2}=0.27228|J|$, respectively.
In the intermediate region, the angle $\vartheta(T)$ is a continuous and smooth
function that monotonically changes from zero to the maximum possible value $\pi/2$.
The square of the optimal measurement angle $\vartheta^2(T)$ varies linearly in the
region $0\le(T_0-T)/T_0\ll1$, as can be seen from Fig.~\ref{fig:zt209B17}, bottom.
Thus,
\begin{equation}
   \label{eq:vartheta}
   \vartheta(T)=
	 \begin{cases}
	    A\!\cdot\!(T_0-T)^\beta,\quad  {\rm if}\ T<T_0 \\
	    0,\qquad\qquad\qquad\!  {\rm if}\ T\ge T_0,
   \end{cases}
\end{equation}
where the critical exponent $\beta=1/2$ and the singularity amplitude $A\approx1.37$,
taking into account that $|J|=1$.
This power-law behavior belongs to the mean-field universality class of criticality.

When the system further evolves along the path $B/|J|=1.7$, quantum correlation
$\Delta$ undergoes a new sudden transition, similar to a second-order phase
transition, at the temperature $T_{\pi/2}=0.27228|J|$.
Post-measurement entropy is approximated near this point by expansion
\begin{eqnarray}
   \label{eq:S_thet1}
   \tilde S(\theta; T,B)&=&\tilde S_{\pi/2}(T,B)
	 +\frac{1}{2!}\tilde S^{\prime\prime}_{\pi/2}(T,B)\!\cdot\!(\theta-\pi/2)^2
   \nonumber\\
	 &&+\frac{1}{4!}\tilde S^{\prime\prime\prime\prime}_{\pi/2}(T,B)\!\cdot\!(\theta-\pi/2)^4+\ldots,
\end{eqnarray}
where $\tilde S_{\pi/2}(T,B)=\tilde S(\pi/2;T,B)$,
$\tilde S^{\prime\prime}_{\pi/2}(T,B)$ equals (\ref{eq:Sii1}), and
$\tilde S^{\prime\prime\prime\prime}_{\pi/2}(T,B)$ is the fourth-order derivative of
$\tilde S(\theta)$ at $\theta=\pi/2$.
Here, the inner minimum reaches a maximum at the endpoint $\theta=\pi/2$ and merges
with him.
The optimal measurement angle will then remain constant at $\pi/2$, up to absolute
zero temperature; see Fig.~\ref{fig:zt209B17}, top.

\begin{figure}
\begin{center}
\epsfig{file=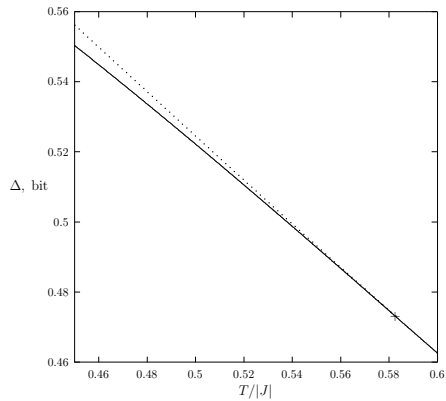,width=5.8cm}
\caption{
Optimized work deficit $\rm\Delta$ (solid line) and branch $\Delta_0$ (dotted line)
upon temperature by  $J_z/|J|=-0.9$ and $B/|J|=1.7$.
The symbol ``plus'' has coordinates $(0.58264,047309)$ and marks the split point of
these two curves.
}
\label{fig:zd209B17}
\end{center}
\end{figure}

Further, the temperature behavior of the optimized one-way quantum work deficit
${\rm\Delta}_\vartheta(T,B)$ near the critical temperature $T_0$ is shown in
Fig.~\ref{fig:zd209B17} by solid line.
%
%
The dependance ${\rm\Delta}_0(T,B)$ drawn with a dotted line is shown for comparison.
Both curves exactly coincide in the temperature region $T\ge T_0=0.58264$.
However the behavior of these functions becomes different for $T<T_0$ -- after the
point marked with ``+'' in Fig.~\ref{fig:zd209B17}.
The error introduced by the ${\rm\Delta}_0$ reaches $1.04\%$ at $T/|J|=0.45$.

\begin{figure}
\begin{center}
\epsfig{file=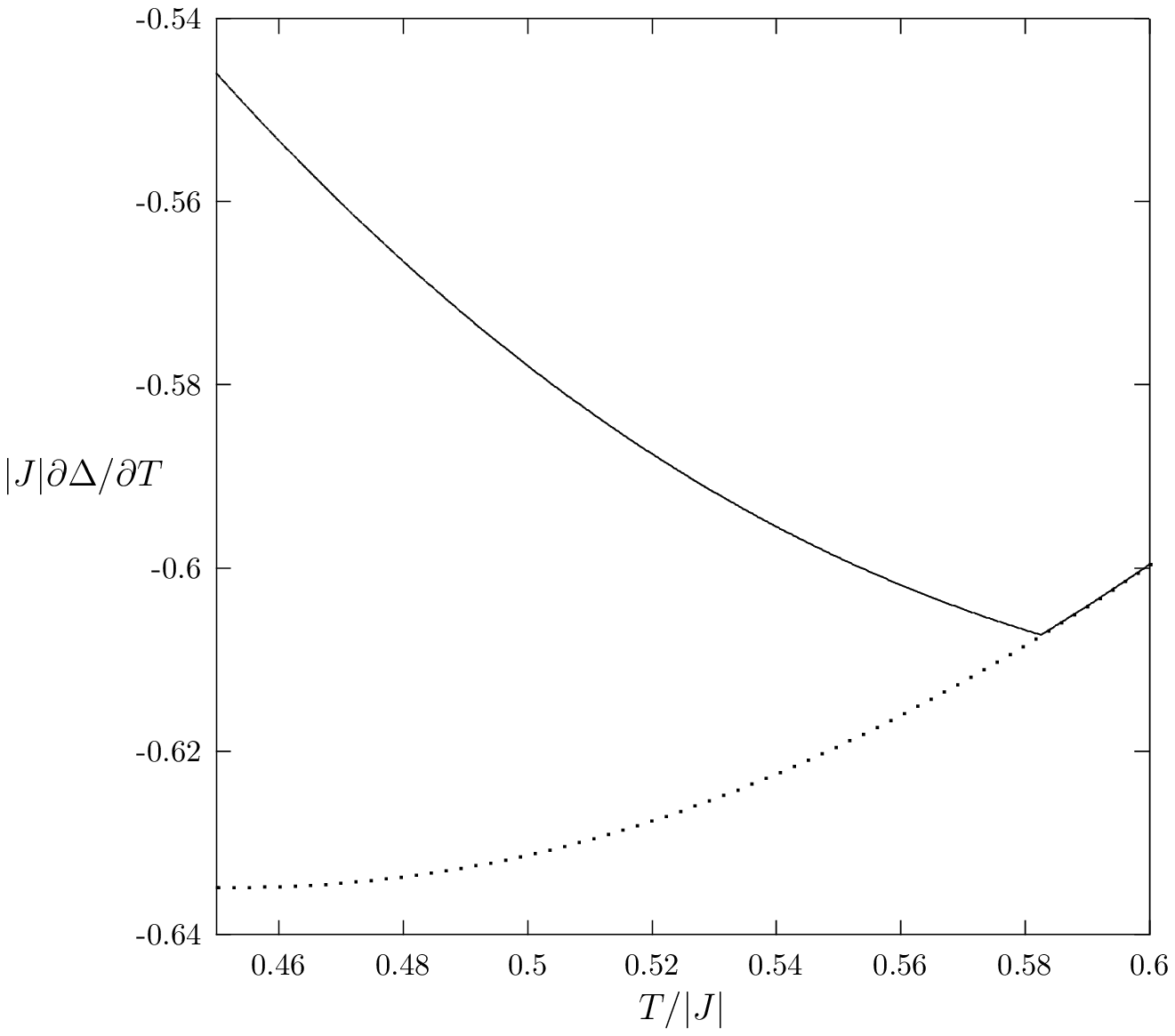,width=5.8cm}
\epsfig{file=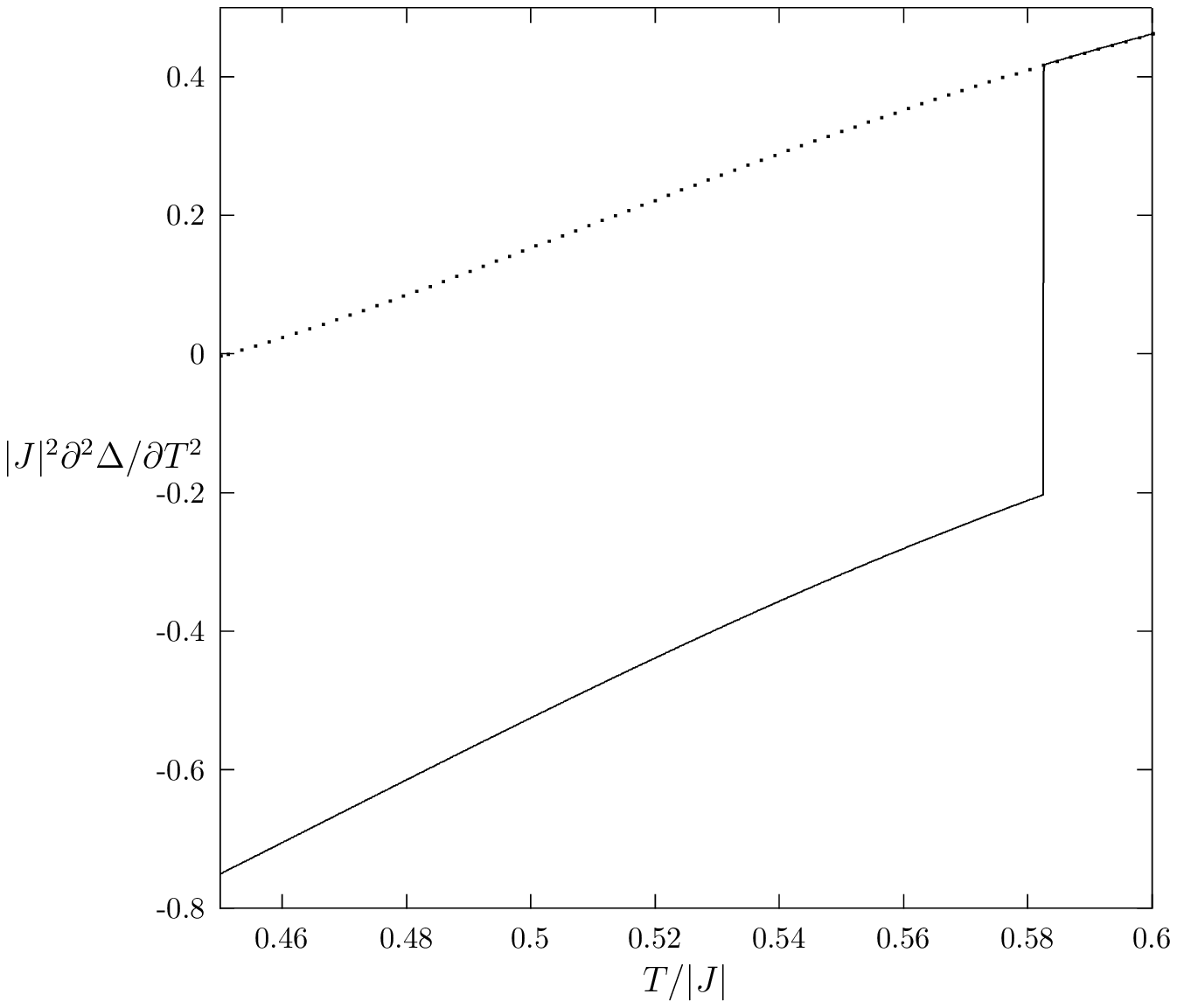,width=5.8cm}
\caption{
The first (top) and second (bottom) derivatives of the optimized work deficit
$\rm\Delta$ (solid lines) and branch ${\rm\Delta}_0$ (dotted lines) depending on
temperature by  $J_z/|J|=-0.9$ and $B/|J|=1.7$.
}
\label{fig:zd12D09B17}
\end{center}
\end{figure}

Dependence of $\rm\Delta$ on temperature in Fig.~\ref{fig:zd209B17} looks like a
continuous and smooth function.
However, this picture breaks down when going to derivatives.
The first derivative of $\rm\Delta$ with respect to temperature is continuous but
exhibits a fracture at $T=T_0$ (see the solid line in Fig.~\ref{fig:zd12D09B17}, top).
%
%
From this we can conclude that the work deficit refers to the functions of
differentiability class $C^1$. 
At the same time, the derivative of the function ${\rm\Delta}_0(T)$ (dotted line)
remains continuous and smooth.  
Moreover, the second derivative of $\rm\Delta$ has discontinuity at $T=T_0$ (solid
line in Fig.~\ref{fig:zd12D09B17}, bottom) and therefore the behavior of work deficit 
here is similar to the behavior in a second-order phase transition.
This is in complete analogy with the Landau theory \cite{LL_StPh,KSG98,A92} and
Ehrenfest classification of phase transitions \cite{J98}.

\subsection{
Discontinuous-continuous combined transitions
}
\label{subsect:1st_order}
It is remarkable that the post-measurement entropy as a function of the measurement
angle may show more complex behavior than unimodal.
Namely, it can have two extrema (one minimum and one maximum) inside the open interval
$(0,\pi/2)$, i.e. exhibit bimodal dependence \cite{Y18}.
This behavior is described by the sixth-order Tailor expansion
\begin{equation}
   \label{eq:S_thet6a}
   \tilde S(\theta; T,B)-\tilde S_0(T,B)\propto\alpha_1\theta^2+\alpha_2\theta^4+\theta^6,
\end{equation}
where $\alpha_i$ ($i=1,2$) are the reduced expansion coefficients.
This leads to a qualitatively new type of sudden transitions.

\begin{figure*}
\begin{center}
\epsfig{file=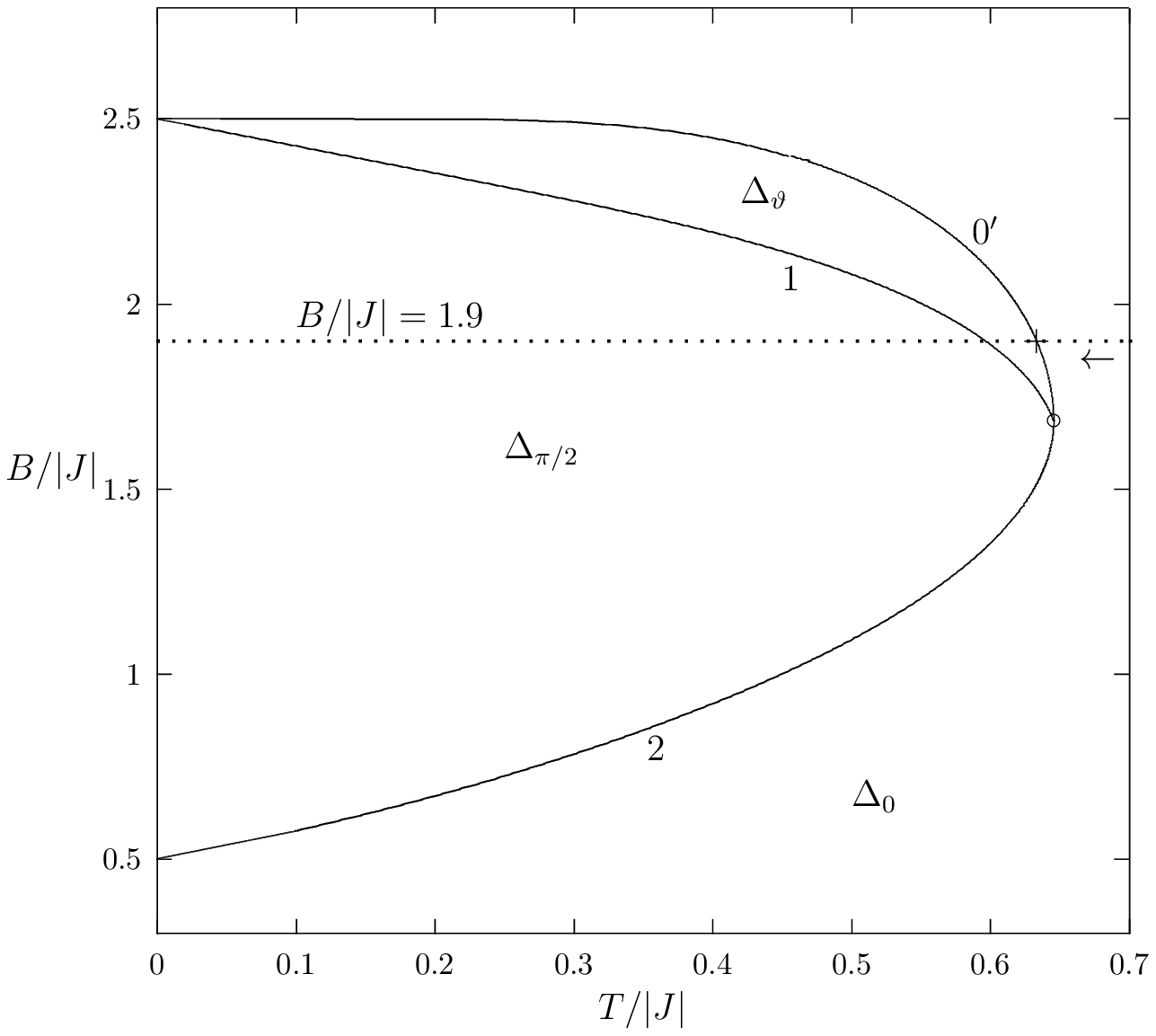,width=7.5cm}
\hspace{5mm}
\epsfig{file=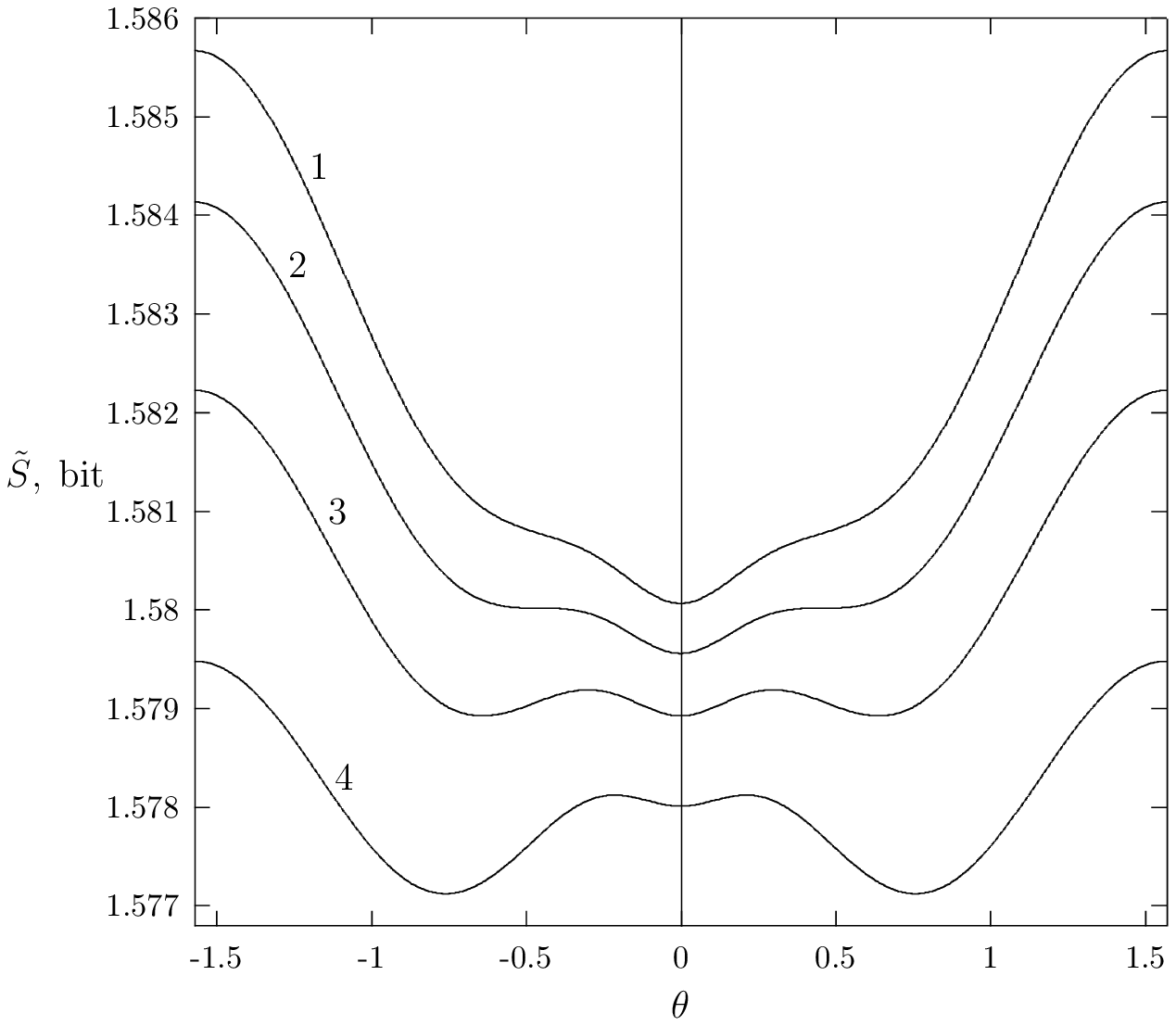,width=7.5cm}
\caption{
(left panel)
Temperature-field phase diagram of work deficit for the model with $J_z/|J|=-1.5$.
Solid lines $0^\prime$ and 1 are the $0^\prime$- and $\pi/2$-boundaries,
respectively; solid line 2 is the boundary between the phases ${\rm\Delta}_0$ and
${\rm\Delta}_{\pi/2}$.
Dotted straight line is the path of system evolution.
The arrow shows the direction of evolution along the path.
The ``+'' symbol has coordinates $(0.63329,1.9)$ and marks an intersection point of
the path and $0^\prime$-boundary.
(right panel)
Post-measurement entropy $\tilde S$ vs measurement angle $\theta$ for $B/|J|=1.9$ and
$T/|J|=0.64~(1), 0.637~(2), 0.63329~(3)$, and $0.628$~(4).
}
\label{fig:zpdzs1way1}
\end{center}
\end{figure*}

\begin{figure}[t]
\begin{center}
\epsfig{file=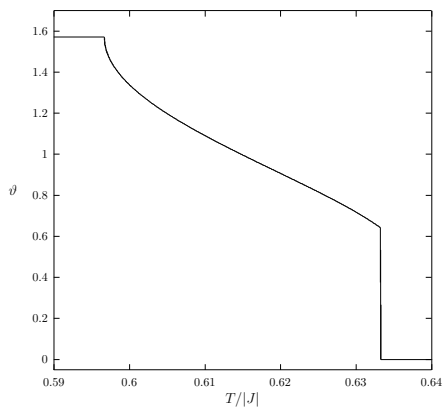,width=5.8cm}
\caption{
Optimal measurement angle $\vartheta$ versus temperature for the coupling
$J_z/|J|=-1.5$ and external field $B/|J|=1.9$.
The jump $\Delta\vartheta=0.64026\approx36.7^\circ$ occurs at the temperature
$T_{c,0}=0.63329|J|$.
}
\label{fig:zt15B19}
\end{center}
\end{figure}
%

Consider the model with coupling $J_z/|J|=-1.5$.
The corresponding phase diagram is shown in Fig.~\ref{fig:zpdzs1way1}, left.
The boundary lines here do not go to infinity, which is consistent with
prediction (\ref{eq:B_as}).

Let the system evolve along the trajectory $B/|J|=1.9$ towards lower temperatures.
In this case, the shape of post-measurement entropy will be deformed, as shown in
Fig.~\ref{fig:zpdzs1way1}, right.
(In order to improve clarity, this figure is presented in a wider window
$[-\pi/2,\pi/2]$.)
At a sufficiently high temperature, the dependence $\tilde S(\theta)$ is
a monotonically increasing function in the interval $ [0,\pi/2] $ (curve~1).
However, at about a temperature $T=0.637|J|$ a pair of local extrema is born inside
the open interval $(0,\pi/2)$. 
This moment is fixed by the curve~2 in Fig.~\ref{fig:zpdzs1way1}, right.
With a further decrease in temperature, the minimum of this pair goes down and at
$T_{c,0}=0.63329|J|$ it reaches the level of post-measurement entropy at $\theta=0$
(curve~3).
At this critical temperature, the optimal measurement angle $\vartheta$ suddenly
switches from zero to the second minimum at $\vartheta=0.64026\approx36.7^\circ$;
see Fig.~\ref{fig:zt15B19}.
The value of this new optimal measurement angle $\vartheta$ is found from the equation
(see Refs.~\cite{Y18,Y19})
\begin{equation}
   \label{eq:0prime}
   \tilde S_0(T,B)=\tilde S_\vartheta(T,B). 
\end{equation}
This equation defines the boundary $0^\prime$, which is shown in 
Fig.~\ref{fig:zpdzs1way1}, left.
As the system cools further, the optimal measurement angle continuously increases
towards $\pi/2$ as shown by curve~4 in Fig.~\ref{fig:zpdzs1way1}, right (see also
Fig.~\ref{fig:zt15B19}).
This limit value of $\vartheta$ is reached at the temperature $T_{c,\pi/2}=0.59669|J|$
where a new sudden transition occurs, now similar to the second order.

\begin{figure}
\begin{center}
\epsfig{file=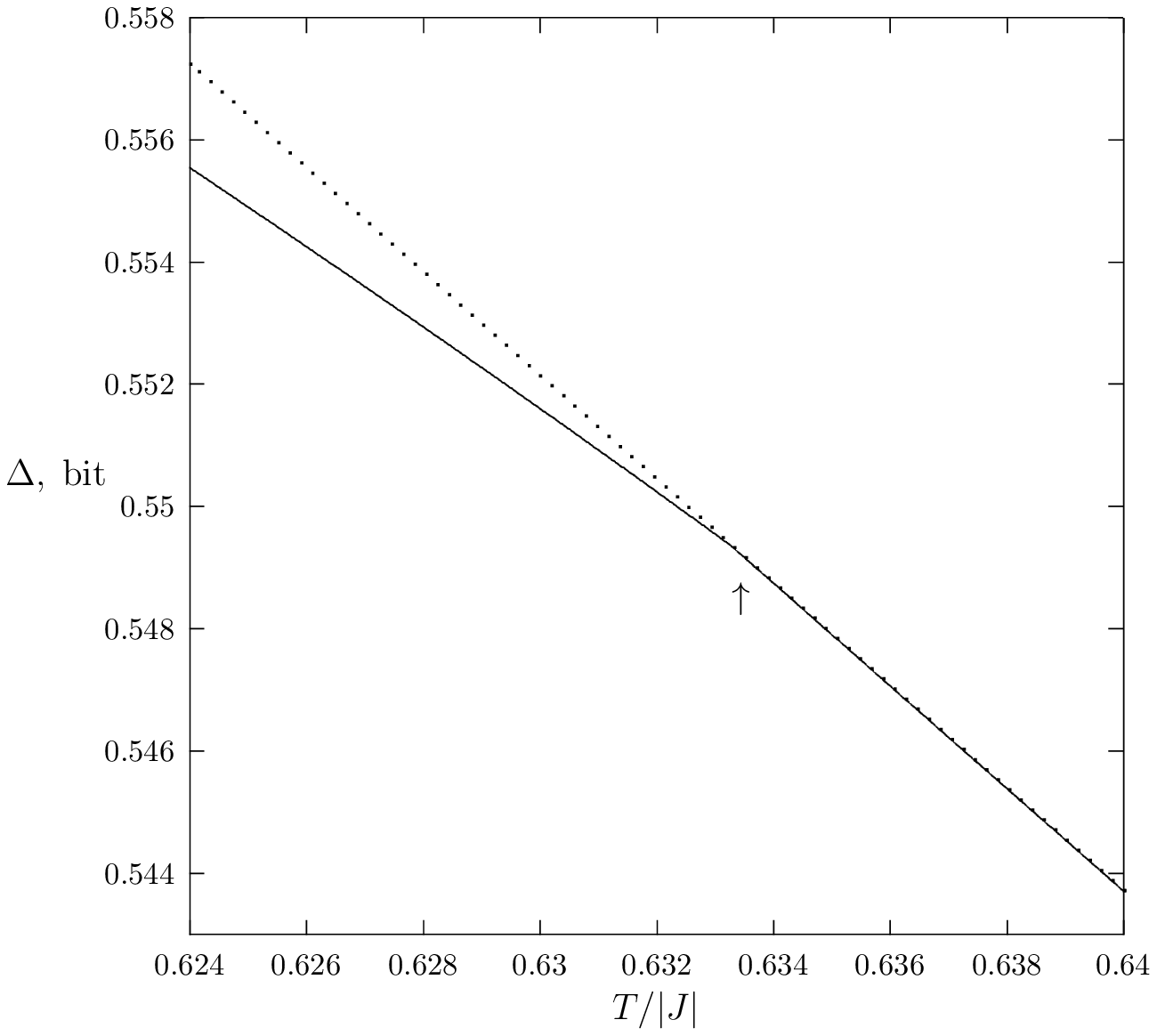,width=5.8cm}
\epsfig{file=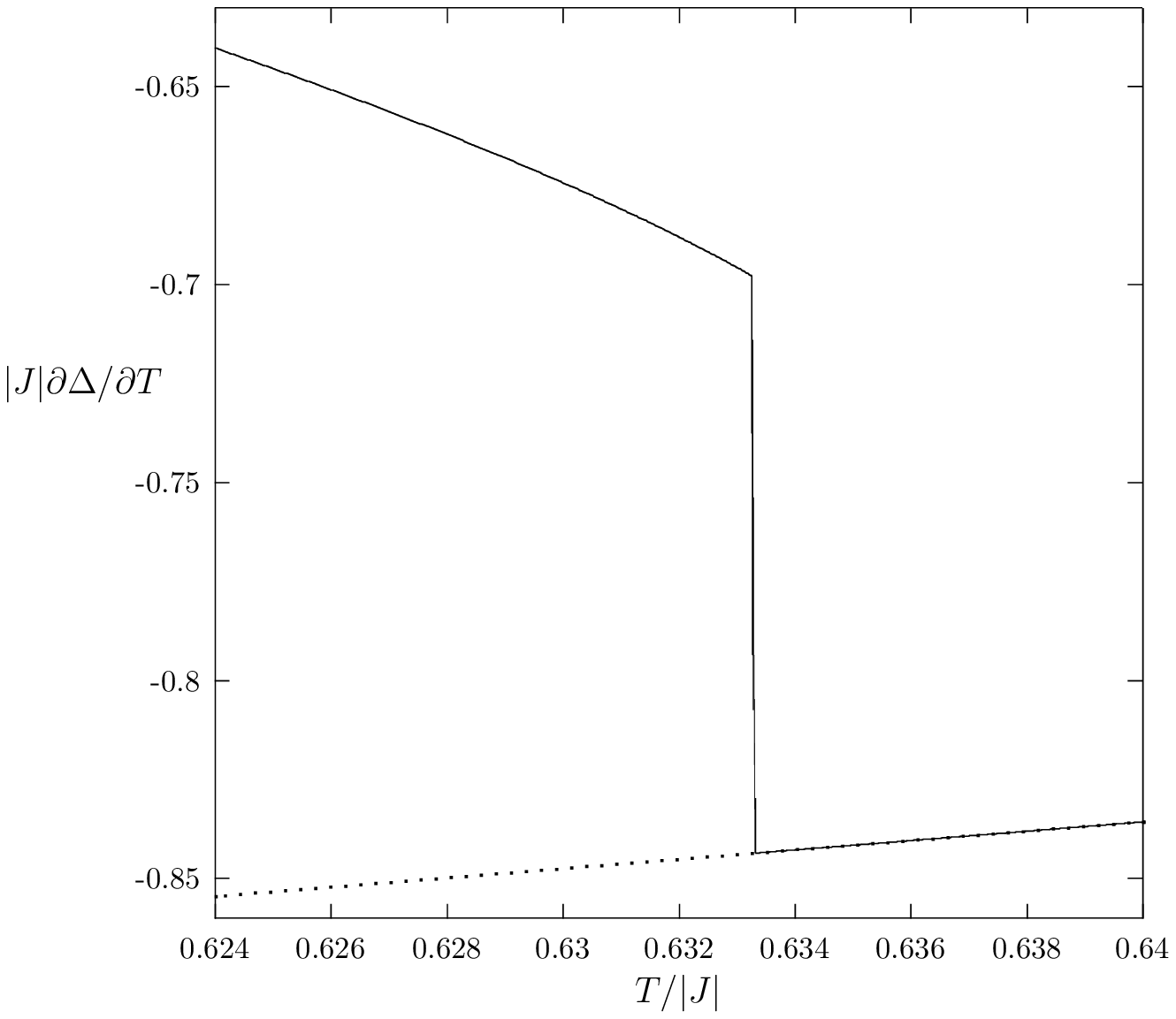,width=5.8cm}
\caption{
Optimal one-way work deficit $\rm\Delta$ (top) and its temperature derivative (bottom)
versus $T/|J|$ by $J_z/|J|=-1.5$ and $B/|J|=1.9$.
The vertical arrow points the position of the fracture on the work deficit curve
at $T/|J|=0.63329$.
The dotted lines correspond to the ${\rm\Delta}_0$ branch and are shown for
comparison.
}
\label{fig:zd1D15B19}
\end{center}
\end{figure}

The dependence ${\rm\Delta}(T)$ is still continuous, but now has a fracture at the
critical temperature $T_{c,0}=0.63329|J|$.
Its behavior is depicted in Fig.~\ref{fig:zd1D15B19}, top.
As a result, the first derivative of the work deficit with respect to temperature
undergoes a finite jump at the critical point, as shown in Fig.~\ref{fig:zd1D15B19},
bottom.
Thus, the described transition combines discontinuous and continuous behavior and is
similar to the first-order phase transitions \cite{KSG98,B87}.

\section{
Continuous transitions of quantum discord
}
\label{sect:QD}
The quantum discord (\ref{eq:Q}) for the system under study is written as
\begin{equation}
   \label{eq:QD}
   Q(T,B)=\min_\theta{\cal Q}(\theta;T,B),
\end{equation}
where ${\cal Q}(\theta;T,B)$ is an non-optimized discord, which is expressed in terms
of Eqs.~(\ref{eq:Spre}) and (\ref{eq:S-postS}) as follows
\begin{eqnarray}
   \label{eq:QD1}
   {\cal Q}(\theta;T,B)&=&\bar S(\theta;T,B)-S(T,B)-(a+b)\ln(a+b)
   \nonumber\\
	 &&-(b+d)\ln(b+d).
\end{eqnarray}
It is clear that this quantity is related with non-optimized work deficit
(\ref{eq:Del_th}) by the equation
\begin{eqnarray}
   \label{eq:QD1a}
   &&{\cal Q}(\theta;T,B)=\Delta(\theta;T,B)-(a+b)\ln(a+b)
   \nonumber\\
	 &&-(b+d)\ln(b+d)
	 +\frac{1+(a-d)\cos\theta}{2}\ln\frac{1+(a-d)\cos\theta}{2}
   \nonumber\\
	 &&+\frac{1-(a-d)\cos\theta}{2}\ln\frac{1-(a-d)\cos\theta}{2}.
\end{eqnarray}
Using the given equation it is easy to establish that quantum discord
at zero optimal measurement angle, $Q_0(T,B)\equiv{\cal Q}(0;T,B)$, equals
${\rm\Delta}_0(T,B)$ and can be written as
\begin{eqnarray}
   \label{eq:Q0}
   Q_0(T,B)&=&\Biggl[\frac{J}{T}\tanh\frac{J}{T}-\ln\Biggl(\cosh\frac{J}{T}\Biggr)\Biggr]
   \nonumber\\
   &&\times\Biggl(1+e^{J_z/T}\cosh\frac{B}{T}\,{\rm sech}\frac{J}{T}\Biggr)^{-1}.
\end{eqnarray}
The high-temperature expansion of this function is equivalent to the right-hand side
of Eq.~(\ref{eq:Del0as1}).

Similarly, the quantum discord at the second endpoint $\theta=\pi/2$ is given as
\begin{eqnarray}
   \label{eq:Q1-a}
   &&Q_{\pi/2}(T,B)=a\ln a+d\ln d+(b+v)\ln(b+v)
   \nonumber\\
	 &&+(b-v)\ln(b-v)
	 -(a+b)\ln(a+b)-(b+d)\ln(b+d)
   \nonumber\\
	 &&-\frac{1+r}{2}\ln\frac{1+r}{2}-\frac{1-r}{2}\ln\frac{1-r}{2},
\end{eqnarray}
where $r=\sqrt{(a-d)^2+4v^2}$ equals (\ref{eq:r}) and $a$, $b$, and $d$ are defined
by Eq.~(\ref{eq:abdv}).
High-temperature behavior is now follow to 
\begin{equation}
   \label{eq:Q1as}
   Q_{\pi/2}(T,B)|_{T\to\infty}=\frac{J^2+J_z^2}{8T^2}-\frac{J^2J_z}{8T^3}+O(1/T^4).
\end{equation}

Next, the second derivative of average conditional entropy with respect to the
measurement angle at $\theta=0$ is equal to
\begin{widetext}
\begin{eqnarray}
   \label{eq:ScondII0}
   {\bar S}_0^{\prime\prime}(T,B)&=&\frac{1}{Z}\bigg\lbrack\!\bigg\lbrack\bigg\lbrace\biggl(\frac{B}{T}
	 +\ln\frac{\exp\bigl((J_z-B)/T)\bigr)+\cosh(J/T)}{\exp\bigl((J_z+B)/T)\bigr)+\cosh(J/T)}\biggr)\sinh\frac{B}{T}
	 \nonumber\\
	 &&+\biggl(\cosh\frac{B}{T}-e^{-J_z/T}\cosh\frac{J}{T}\biggr)\bigg\lbrack\frac{J_z}{T}-\ln\biggl(\cosh\frac{J}{T}\biggr)\bigg\rbrack\bigg\rbrace e^{J_z/2T}
	 \nonumber\\
   &&-\frac{1}{2}e^{-J_z/2T}\bigg\lbrack\frac{(J_z+B)/T-\ln\bigl(\cosh(J/T)\bigr)}{\exp\bigl((J_z+B)/T\bigr)-\cosh(J/B)}
	 +\frac{(J_z-B)/T-\ln\bigl(\cosh(J/T)\bigr)}{\exp\bigl((J_z-B)/T\bigr)-\cosh(J/B)}\bigg\rbrack\sinh^2\frac{J}{T}\bigg\rbrack\!\bigg\rbrack.
\end{eqnarray}
\end{widetext}
Its high-temperature expansion is given as
\begin{eqnarray}
   \label{eq:hTScondII0}
   &&{\bar S}_0^{\prime\prime}(T,B)|_{T\to\infty}=
   \frac{J_z^2-J^2}{4T^2}
	 \nonumber\\
	 &&+\frac{5J^4+2B^2J^2-3B^2J_z^2-4J_z^2J^2-J_z^4}{48T^4}+O(1/T^5).
	 \nonumber\\
\end{eqnarray}
In similar way,
\begin{eqnarray}
   \label{eq:S2primepi2a}
	 \bar S_{\pi/2}^{\prime\prime}(T,B)
	 &=&2\frac{v^2}{r}\bigg[1-\bigg(\frac{a-2b+d}{r}\bigg)^2\bigg]\ln\frac{1+r}{1-r}
	 \nonumber\\
	 &&\frac{1}{2}(a-d)^2\biggl[2-\frac{1}{1+r}\bigg(1+\frac{a-2b+d}{r}\biggr)^2
	 \nonumber\\
	 &&-\frac{1}{1-r}\bigg(1-\frac{a-2b+d}{r}\biggr)^2\biggr]
\end{eqnarray}
and its asymptotics at high temperatures is given as
\begin{equation}
   \label{eq:hTScondII1}
   {\bar S}_{\pi/2}^{\prime\prime}(T,B)|_{T\to\infty}=\frac{J^2-J_z^2}{4T^2}+O(1/T^4).
\end{equation}

Equations for the 0- and $\pi/2$-boundaries are similar to (\ref{eq:S11}) and written
as
\begin{equation}
   \label{eq:barS11}
   \bar S^{\prime\prime}_0(T,B)=0\quad
   {\rm and}\quad
   \bar S^{\prime\prime}_{\pi/2}(T,B)=0.
\end{equation}
Together with Eqs.~(\ref{eq:ScondII0}) and (\ref{eq:S2primepi2a}) they allow to find the phase diagram of quantum discord.

\begin{figure}
\begin{center}
\epsfig{file=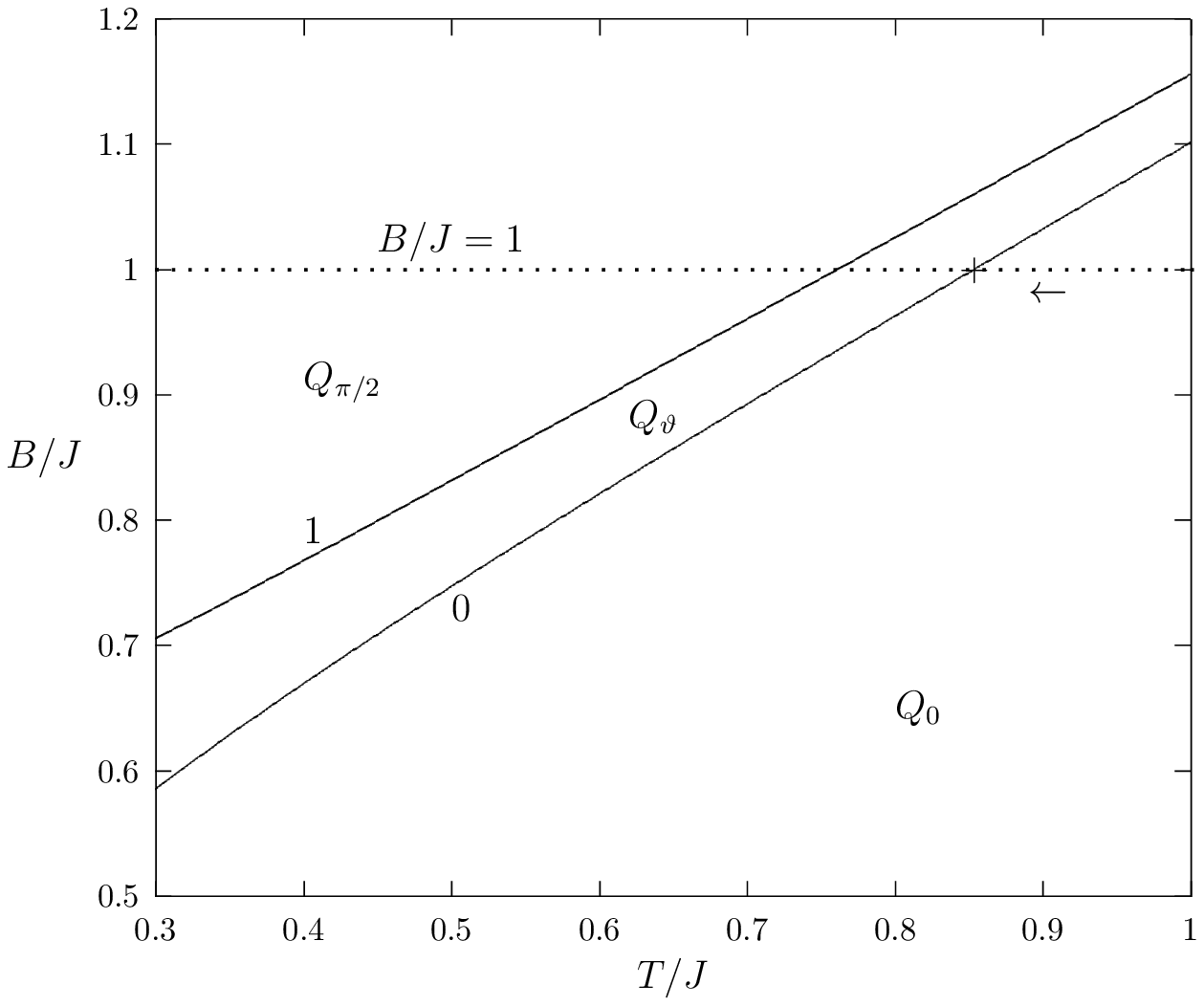,width=7.1cm}
\hspace{5mm}
\epsfig{file=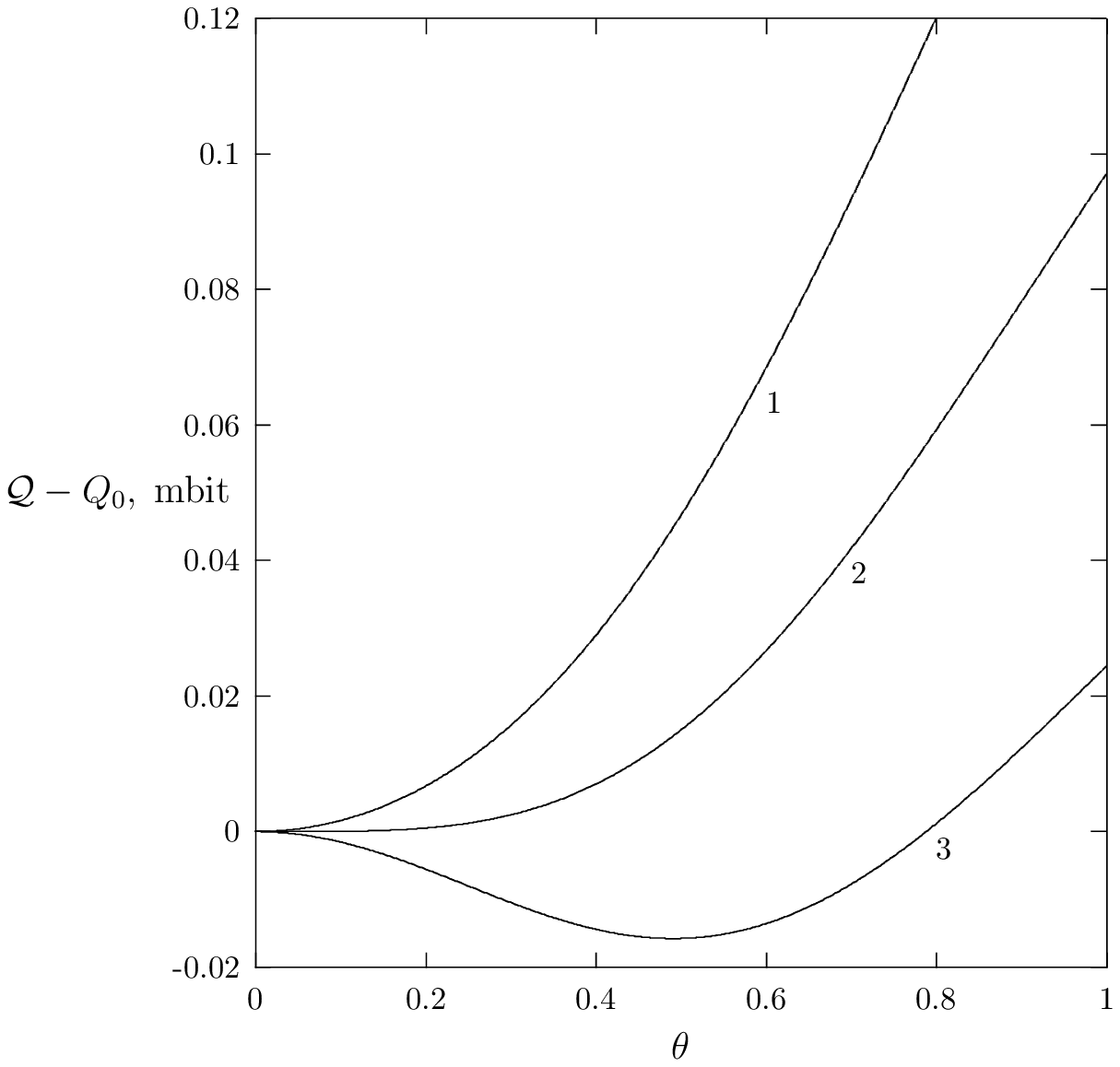,width=7.1cm}
\caption{
(top panel)
A fragment of $(T,B)$ phase diagram for the quantum discord in the model with
interaction constants $J=1$ and $J_z=1.02$.
Solid lines 0 and 1 are the 0- and $\pi/2$-boundaries, respectively.
Dotted straight line is the trajectory of the system evolution, and arrow shows
direction for such evolution.
Symbol ``+'' has coordinates (0.85361,1) and marks the critical point.
(bottom panel)
Variation of non-optimized discord ${\cal Q}-Q_0$ versus measurement angle
$\theta\in[0,1]$ for temperatures $T/J=0.88$~(1), 0.85361~(2), and 0.83~(3);
1~mbit$=1\times10^{-3}$~bit.
}
\label{fig:z102a}
\end{center}
\end{figure}

If the optimal measurement angle $\vartheta$ changes continuously from zero to
$\pi/2$, then this will be similar to the continuous (second-order) phase transitions
of the Landau theory.
It is remarkable that such situations do occur in the system under study.
Indeed, consider the model with interaction constants $J=1$ and $J_z=1.02$.
Numerically solving the above equations (\ref{eq:barS11}), we find the phase diagram
shown in Fig.~\ref{fig:z102a}, top.
Let the system evolve with a change in temperature along the path $B/J=1$.
During this evolution, the conditional entropy and therefore the non-optimized
discord (which presented minus the constant part $Q_0$) qualitatively change their
behavior, see Fig.~\ref{fig:z102a}, bottom.
The function ${\cal Q}(\theta)$ monotonically increases at sufficiently high
temperatures (curve~1 in Fig.~\ref{fig:z102a}, bottom) and its behavior near
$\theta=0$ is similar to $\sim\theta^2$.
However, the behavior of this dependence changes qualitatively with approaching the
0-boundary (see Fig.~\ref{fig:z102a}, top) -- its shape converts to $\sim\theta^4$ at
the Curie-like point $T_{C, 0}= 0.85361J$ (curve~2 in Fig.~\ref{fig:z102a}, bottom).
Then, upon further cooling of the spin system, the minimum continuously shifts inside
interval $(0,\pi/2)$; curve~3 in Fig.~\ref{fig:z102a}, bottom.
This behavior near the point $T_{C,0}$ is well approximated by Landau-like expansion
\begin{equation}
   \label{eq:Q_thet}
   {\cal Q}(\theta; T,B)-Q_0(T,B)\approx
	 +\frac{1}{2!}\bar S^{\prime\prime}_0(T,B)\!\cdot\!\theta^2
	 +\frac{1}{4!}\bar S^{\prime\prime\prime\prime}_0(T,B)\!\cdot\!\theta^4.
\end{equation}
Here $Q_0(T,B)$ and $\bar S^{\prime\prime}_0(T,B)$ are given by Eqs.~(\ref{eq:Q0}) and
(\ref{eq:ScondII0}), respectively; expression for the fourth derivative,
$\bar S^{\prime\prime\prime\prime}_0(T,B)$, is too cumbersome and therefore we do not
present it.

\begin{figure}[t]
\begin{center}
\epsfig{file=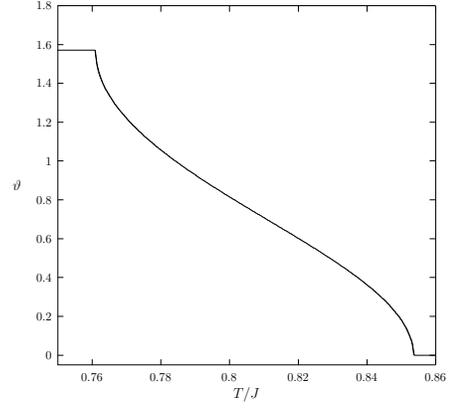,width=5.8cm}
\caption{
Optimal measurement angle $\vartheta$ as a function of
temperature by $J=1$, $J_z=1.02$, and $B/J=1$.
The Curie-like points are $T_{C,\pi/2}/J=0.76106$ and $T_{C,0}/J=0.85361$.
}
\label{fig:zt102}
\end{center}
\end{figure}

With decreasing temperature, the interior minimum tends continuously to the value
$\theta=\pi/2$, which is attained at temperature $T_{C,\pi/2}=0.76106J$, as depicted
in Fig.~\ref{fig:zt102}.
%
\begin{figure}[t]
\begin{center}
\epsfig{file=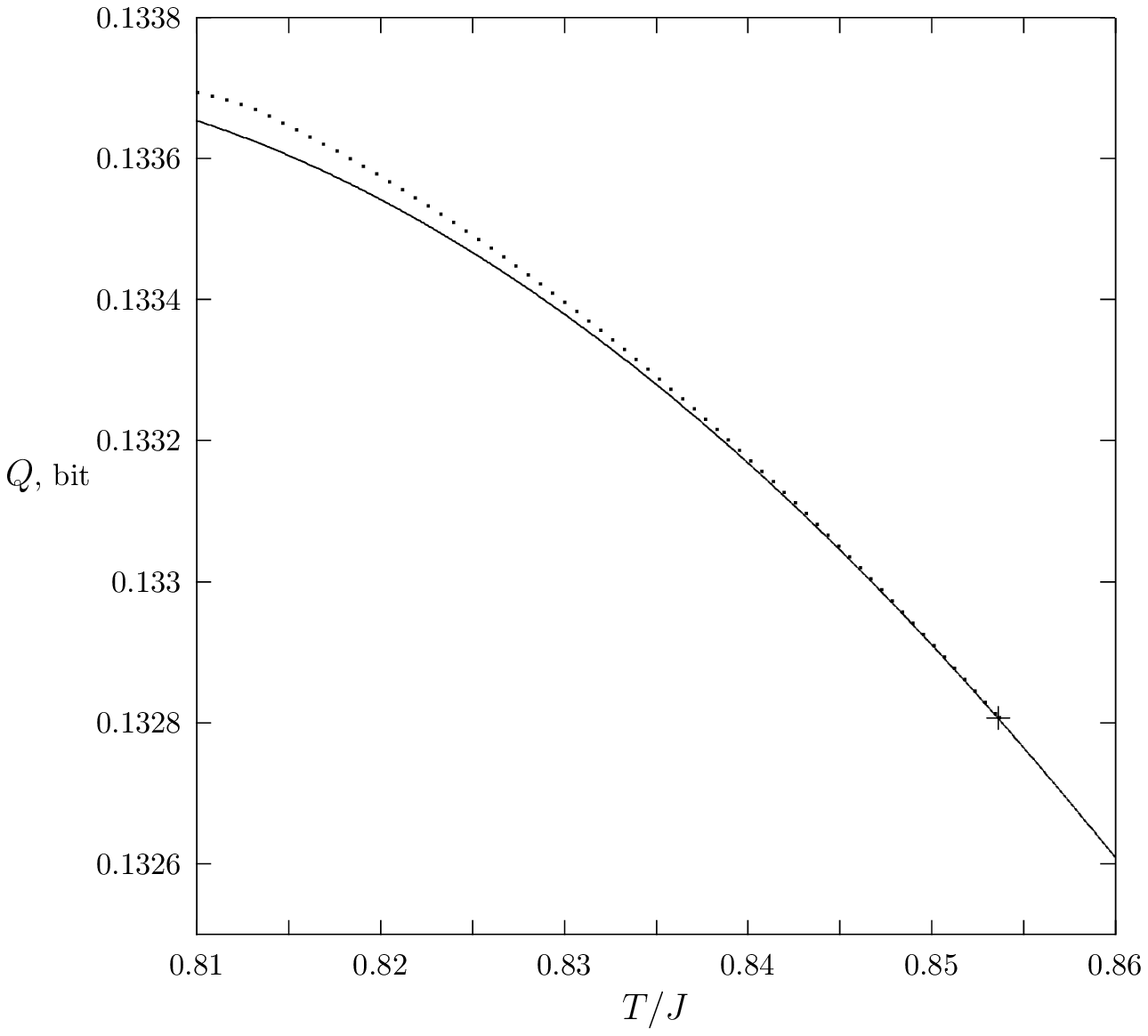,width=5.8cm}\\
\epsfig{file=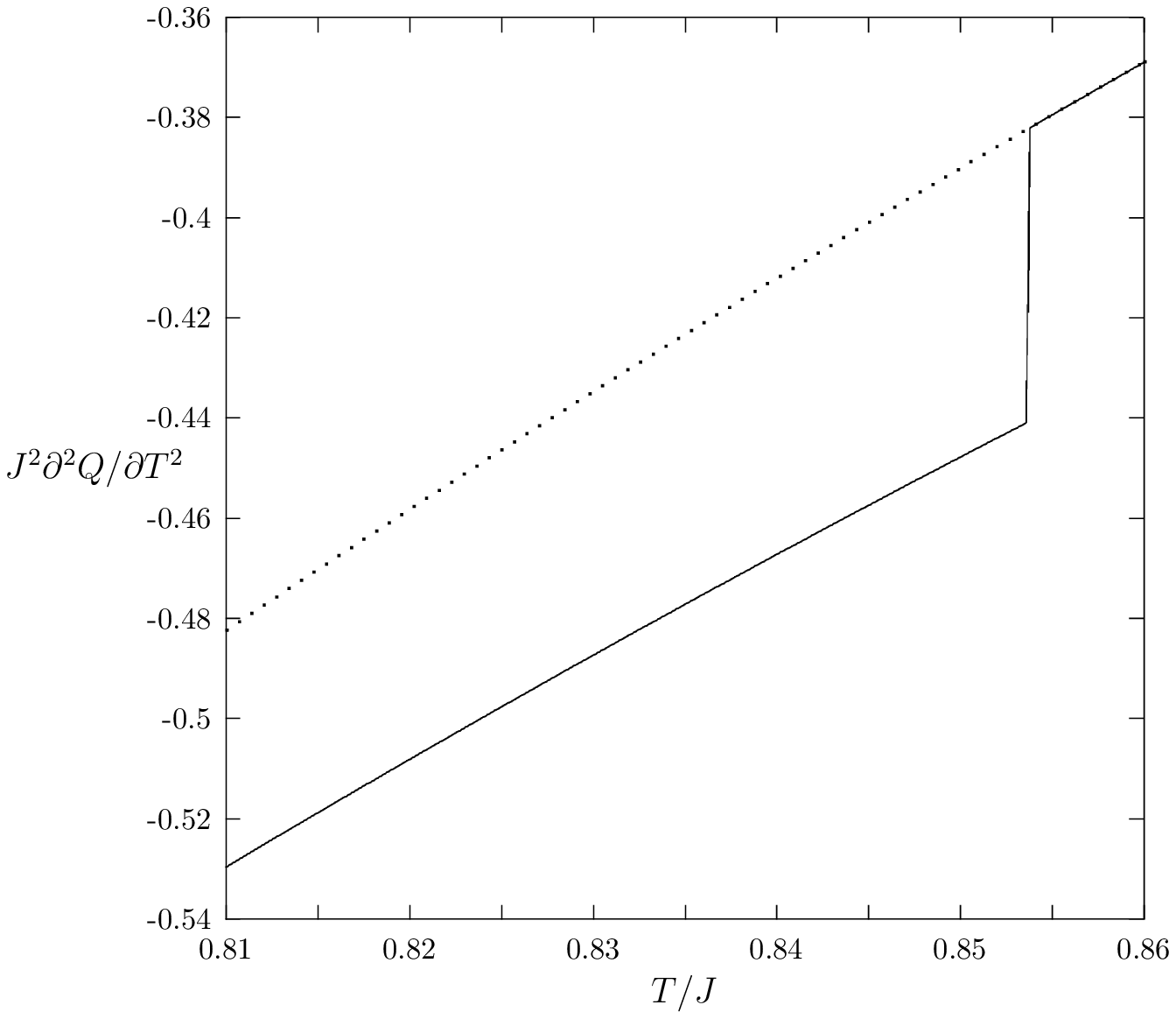,width=5.8cm}
\caption{
Quantum discord $Q$ (top) and its second derivative (bottom) near the Curie-like point
$T_{C,0}=0.85361J$  by $J=1$, $J_z=1.02$, and $B/J=1$.
Temperature dependencies of $Q_0$ (dotted lines) are shown for comparison.
The ``+'' symbol has coordinates $(0.85361,0.13281)$ and
corresponds to the splitting point of the curves $Q(T)$ and $Q_0(T)$.
}
\label{fig:zq102a}
\end{center}
\end{figure}
At this second Curie-type point, another second-order phase transition occurs.
Below the temperature $T_{C,\pi/2}$, the optimal measurement angle again takes
stationary value, now equal to $\pi/2$.
Near the critical point $T_{C,0}$, the value $\vartheta(T)$ follows to
law~(\ref{eq:vartheta}) and around $T_{C,\pi/2}$ changes as
\begin{equation}
   \label{eq:vartheta1}
   \vartheta(T)=
	 \begin{cases}
 	    \pi/2-A\!\cdot\!(T-T_{C,\pi/2})^\beta,\quad  {\rm if}\ T>T_{C,\pi/2} \\
	    \pi/2,\qquad\qquad\qquad\qquad\quad\   {\rm if}\ T\le T_{C,\pi/2},
   \end{cases}
\end{equation}
where the critical exponent $\beta=1/2$.
So, near the critical points $T_{C,0}$ and $T_{C,\pi/2}$, the system experiences
continuous sudden transitions.
Here, quantum discord at any temperature $T\in[0,\infty)$ is a continuous and smooth
function.
However, its second-order derivatives with respect to temperature have finite jumps in
both critical points.
This is illustrated in Fig.~\ref{fig:zq102a} for the Curie-like point $T_{C,0}$.
By analogy with the Ehrenfest classification, we assign these continuous
transitions to sudden transitions/changes of the second order.

\section{
Conclusions
}
\label{sect:Concl}
It is well known that quantum correlations can be used to detect quantum phase
transitions.
For instance, the next-nearest-neighbor entanglement has a maximum at the
critical point in the one-dimensional transverse-field Ising system \cite{ON02},
entanglement shows scaling behavior in the vicinity of the transition point
\cite{OAFF02}, bipartite entanglement indicates the quantum phase transitions in the
frustrated spin models \cite{BB21}, Floquet dynamical phase transition is signaled by
the behavior of entanglement spectrum \cite{JA21}.

To an even greater extent it manifests itself for quantum discord and information
deficit.
Using quantum discord, it was shown that the magnitude of quantum correlations
increases in the region close to the critical points \cite{D08}, and that it exhibits
signatures of quantum phase transitions \cite{S09}.
Moreover, the quantum discord makes it possible to confidently determine the points of
the quantum phase transition at {\em finite} temperatures \cite{WTRR10,WRR11}.
(Using available solution for the XXZ chain in an external longitudinal field,
the authors \cite{WRR11} also concluded, adding, however, numerical arguments, that
the quantum discord of the two-site reduced density matrix does not contain regions
with the variable optimal measurement angle; the question for the one-way quantum work
deficit remains open.)

On the other hand, quantum correlations are considered as a physical resource (``as
real as energy'' \cite{HHHH09}) and they themselves can undergo sudden changes.
In Refs.~\cite{MCSV09,LC10,CM17}, the sudden changes such as simple jumps (switches)
from one stationary value of the optimal parameter to another without its continuous
shift were studied in two-qubit systems.

In the present paper, we have shown that at the boundaries of regions with variable
and constant optimal measurement angles, quantum discord exhibits a sudden change of
second-order, while one-way quantum work deficit can experience sudden changes of both
the first and second kinds.
These phenomena shed additional light on the properties of nonclassical correlations
in physical systems.

Our consideration is limited to the temperature dependencies of quantum correlations.
It would be interesting to analyze similar dependencies on other control parameters,
such as $J$, $B$, and $J_z$.
These studies will probably require the use of catastrophe theory methods \cite{A92}.

\section*{ACKNOWLEDGMENT}
This work was supported by the program
CITIS \#AAAA-A19-119071190017-7.



\end{document}